\documentclass[prd,twocolumn,showpacs,nofootinbib,floatfix,amsmath,amssymb,floatfix]{revtex4}
\usepackage{graphicx,color,dcolumn,booktabs,bm}
\usepackage{longtable,lscape}
\usepackage{txfonts}
\usepackage{overpic}
\usepackage{amssymb}
\usepackage{indentfirst}
\usepackage{feynmf}   
\usepackage{slashed}  
\usepackage{cases}
\usepackage{color}
\usepackage{multirow}
\usepackage{graphicx,color,dcolumn,booktabs,bm}
\usepackage{epsfig,dsfont,amssymb,amsmath,amsfonts,amsbsy,mathrsfs}

\graphicspath{{Figures/}} %
\usepackage[colorlinks, citecolor=blue,anchorcolor=red,menucolor=red, linkcolor=red,filecolor=red,runcolor=red,urlcolor=blue,frenchlinks=red]{hyperref}
\begin{document}

\title{Numerical analysis of the production of $D^{(*)}(3000)$, $D_{sJ}(3040)$ and their partners through the semileptonic decays of $B_{(s)}$ mesons in terms of the light-front quark model}
\author{Hao Xu$^{1,2}$}\email{xuh2013@lzu.edu.cn}
\author{Qi Huang$^{1,2}$}\email{huangqi19920518@hotmail.com}
\author{Hong-Wei Ke$^3$}\email{khw020056@hotmail.com}
\author{Xiang Liu$^{1,2}$\footnote{Corresponding author}}\email{xiangliu@lzu.edu.cn}
\affiliation{$^1$School of Physical Science and Technology, Lanzhou University,
Lanzhou 730000, China
\\$^2$Research Center for Hadron and CSR Physics,
Lanzhou University $\&$ Institute of Modern Physics of CAS,
Lanzhou 730000, China\\
$^3$School of Science, Tianjin University, Tianjin, 300072, China}

\begin{abstract}

Inspired by the newly observed $D^{(*)}(3000)$, and $D_{sJ}(3040)$ states, in this work we study the production of $D^{(*)}(3000)$, $D_{sJ}(3040)$ and their partners through the semileptonic decays of $B_{(s)}$ mesons, where the light-front Quark model is applied to the whole calculation. Our numerical results indicate that the $B_{(s)}$ semileptonic decays into the $2P$ states of the charmed or charmed-strange meson family have considerable branching ratios, which shows that these semileptonic decays can be accessible at future experiments, especially LHCb and the forthcoming Belle II.

\end{abstract}

 \maketitle

\section{introduction}\label{sec1}

In recent years, experiments have made progress in searching for higher charmed and charmed-strange mesons, where more and more charmed or charmed-strange states were reported, which has stimulated theorist's interest in revealing their underlying properties (see Ref. \cite{Liu:2010zb} for a brief review).

Among all observed charmed and charmed-strange states, there are three states $D(3000)$, $D^*(3000)$, and $D_{sJ}(3040)$ with masses around 3 GeV. $D(3000)$ and $D^*(3000)$ were observed by the LHCb Collaboration \cite{Aaij:2013sza} by measuring the $D^+\pi^-$, $D^0\pi^+$, and $D^{*+}\pi^-$ invariant mass spectra from the inclusive processes $pp\to D^+\pi^-X$, $pp\to D^0\pi^+X$, and $pp\to D^{*+}\pi^-X$, where $X$ is a system composed of any collection of charged and neutral particles \cite{Aaij:2013sza}.  $D^*(3000)$ appears in the $D^+\pi^-$ invariant mass spectrum, while $D(3000)$ exists in the $D^{*+}\pi^-$
invariant mass spectrum. The resonance parameters of $D(3000)$ and $D^*(3000)$ are
$m_{D(3000)}=2971.8\pm8.7$ MeV, $\Gamma_{D(3000)}=188.1\pm44.8$ MeV,
$m_{D^*(3000)}=3008.1\pm4.0$ MeV, and $\Gamma_{D^*(3000)}=110.5\pm11.5$ MeV.
$D(3000)$ and $D^*(3000)$ can be explained as the $2P$ states in the $D$-meson family
\cite{Sun:2013qca}, i.e., $D(3000)$ and $D^*(3000)$ are the first radial excitations of $D_1(2430)$ ($J^P=1^+$) and $D_0^*(2400)$ ($J^P=0^+$), respectively \cite{Sun:2013qca}, which was also supported by later work \cite{Yu:2014dda}.
Before observing $D(3000)$ and $D^*(3000)$,  the BaBar Collaboration announced the observation of the charmed-strange state $D_{sJ}(3040)$ only in the $D^*K$ invariant mass spectrum in inclusive $e^+e^-$ interactions \cite{Aubert:2009ah} , which has a mass $m_{D{sJ}(3040)}=3044\pm8(\mathrm{stat})^{+30}_{-5}(\mathrm{syst})$ MeV and width $\Gamma_{D{sJ}(3040)}=239\pm35(\mathrm{stat})^{+46}_{-42}(\mathrm{syst})$ MeV. Here, $D_{sJ}(3040)$ is a good candidate for the first radial excitation of $D_{s1}(2460)$, as indicated in Ref. \cite{Sun:2009tg}.

Although there is abundant experimental information of regarding $D(3000)$, $D^*(3000)$, and $D_{sJ}(3040)$, we have noticed that they can also be produced (theoretically) via the semileptonic decays of $B_{(s)}$, which is different from the reported production processes of these states. The semileptonic decays of $B_{(s)}$ can be a good platform to study $D(3000)$, $D^*(3000)$, and $D_{sJ}(3040)$, and previous theoretical efforts have studied the production of newly observed charmed and charmed-strange mesons through these decays. For example, $B_s\to D_{s0}^*(2317)\ell \bar{\nu}_\ell,\,D_{s1}(2460)\ell \bar{\nu}_\ell$ were calculated using QCD Sum rules \cite{Huang:2004et,Aliev:2006qy,Aliev:2006gk}, the constituent quark-meson model \cite{Zhao:2006at}, and light-cone QCD Sum rule \cite{Li:2009wq}. In Ref. \cite{Li:2010bb}, Li {\it et al.} studied
the $B_s\to D_{sJ}(3040)\ell \bar{\nu}_\ell$ semileptonic decays using the covariant light-front quark model. These studies showed that these semileptonic decays have considerable branching ratios.

In this work, we explore the production of $D^{(*)}(3000)$, $D_{sJ}(3040)$, and their partners through the semileptonic decays of $B_{(s)}$ mesons, which is helpful to estimate the discovery potential of these states via these corresponding decays. It is obvious that this information is valuable to future experimental searches for $D^{(*)}(3000)$, $D_{sJ}(3040)$, and their partners via the $B_{(s)}$ semileptonic decays.
As a relativistic quark model, the light-front quark model has been applied to investigate the transitions among mesons, where the obtained results agree with the experimental data within a reasonable error tolerance \cite{Jaus:1989au,Jaus:1991cy,Jaus:1999zv,Ji:1992yf,Cheng:1996if,Cheng:2003sm,Hwang:2006cua,Wang:2007sxa,Lu:2007sg,Choi:2007se,Ke:2009ed,Ke:2010vn,
Li:2010bb,Wei:2009nc,Ke:2009mn,Ke:2011fj,Ke:2011mu,Ke:2011jf,Ke:2013yka,Shen:2008zzb,Shen:2013oua}. Thus, in this work we adopt the light-front quark model to calculate the production of $D^{(*)}(3000)$, $D_{sJ}(3040)$, and their partners through the semileptonic decays of $B_{(s)}$ mesons. In the next sections, we will present the details of the calculation.

This paper is organized as follows. In Sec. \ref{sec2}, we list the hadronic matrix elements and the corresponding form factors. In Sec. \ref{sec3}, the numerical results including the obtained form factors and the decay branching ratios are given. The final section is devoted to a summary of our work.

\section{The hadronic matrix elements and the calculation of the corresponding form factors}\label{sec2}

In this work, we study the production of $D^{(*)}(3000)$, $D_{sJ}(3040)$ and their partners via the semileptonic decays of $B$ and $B_s$ mesons. The effective weak Hamiltonian involved in the $B^+(0^-) \to \bar{D}_J^0$ and $B_s^0(0^-) \to D_{sJ}^-$ transitions is
\begin{eqnarray}
 H_{\rm eff}&=& \frac{G_F}{\sqrt 2} V_{cb}\left [\bar
 c\gamma_\mu(1-\gamma_5)b\right]\left[\bar \ell\gamma^\mu(1-\gamma_5)\nu\right],
\end{eqnarray}
where $G_F$ is the Fermi coupling constant and $V_{cb}$ denotes the Cabibbo-Kobayashi-Maskawa (CKM) matrix element.

The hadronic matrix elements of $B^+(0^-) \to \bar{D}_J^0$ and $B_s^0(0^-) \to D_{sJ}^-$ decays
can be obtained by introducing the form factors, i.e.,
 \begin{eqnarray}
   && \langle V(P^{\prime\prime},\varepsilon^{\prime\prime*})|V_\mu|B_{(s)}(P^\prime)\rangle \nonumber\\&&=
       -\frac{1}{ m_{B_{(s)}}+m_V}\,\epsilon_{\mu\nu\alpha \beta}\varepsilon^{\prime\prime*\nu}P^\alpha
    q^\beta
      V^{B_{(s)}\to V}(q^2), \label{eq:B TO_V1}
      \end{eqnarray}
      \begin{eqnarray}
   &&  \langle V(P^{\prime\prime},\varepsilon^{\prime\prime*})|A_\mu|
     B_{(s)}(P^\prime)\rangle\nonumber\\&&=i\Big\{
         (m_{B_{(s)}}+m_{V})\varepsilon^{\prime\prime*}_\mu A_1^{B_{(s)} \to V}(q^2)
          -\frac{\varepsilon^{\prime\prime*}\cdot P}
         { m_{B_{(s)}}+m_{V}}\,
         P_\mu A_2^{B_{(s)}\to V}(q^2) \nonumber\\ &&
       \quad-2m_{V}\,{\frac{\varepsilon^{\prime\prime*}\cdot P}{
    q^2}}\,q_\mu\big[A_3^{B_{(s)}\to V}(q^2)-A_0^{B_{(s)}\to V}(q^2)\big]\Big\},\label{eq:B TO_V}
    \end{eqnarray}
      \begin{eqnarray}
      && \langle S(P^{\prime\prime})|A_\mu| B_{(s)}(P^\prime)\rangle\nonumber\\&&=-i\Bigg[
        \left(P_\mu-\frac{m_{B_{(s)}}^2-m_S^{2}}{q^2}q_\mu\right) F_1^{B_{(s)}\to S}(q^2)\nonumber\\&&
        \quad+\frac{m_{B_{(s)}}^2-m_S^{2}}{q^2}q_\mu F_0^{B_{(s)}\to S}(q^2)\Bigg],\label{eq:B TO_S}
         \end{eqnarray}
      \begin{eqnarray}
       &&\langle A(P^{\prime\prime},\varepsilon^{\prime\prime*})|A_\mu| B_{(s)}(P^\prime)\rangle\nonumber\\&&\quad=-
     \frac{\epsilon_{\mu\nu\alpha \beta}}{ m_{B_{(s)}}-m_{A}}\,\varepsilon^{\prime\prime*\nu}P^\alpha
    q^\beta  A^{B_{(s)}\to A}(q^2),   \nonumber\\\label{eq:B TO A1}
    \end{eqnarray}
      \begin{eqnarray}
      && \langle A(P^{\prime\prime},\varepsilon^{\prime\prime*})|V_\mu|
     B_{(s)}(P^\prime)\rangle \nonumber\\&&= -i\Big\{
         (m_{B_{(s)}}-m_{A})\varepsilon^{\prime\prime*}_\mu V_1^{B_{(s)}\to A}(q^2) -\frac{\varepsilon^{\prime\prime*}\cdot P}
         { m_{B_{(s)}}-m_{A}}\,
         P_\mu V_2^{B_{(s)}\to A}(q^2)   \nonumber\\ &&
          \quad-2m_{A}\,{\varepsilon^{\prime\prime*}\cdot P\over q^2}\,q_\mu\big[V_3^{B_{(s)}\to A}(q^2) -V_0^{B_{(s)}\to A}(q^2)\big]\Big\},\label{eq:B TO A}
          \end{eqnarray}
      \begin{eqnarray}
       &&  \langle T(P^{\prime\prime},\varepsilon^{\prime\prime*})|V_\mu| B_{(s)}(P^\prime)\rangle
=h(q^2)\epsilon_{\mu\nu\alpha\beta}\varepsilon^{\prime\prime*\nu\lambda}P_\lambda
 P^\alpha q^\beta,\label{eq:B TO T1}
 \end{eqnarray}
      \begin{eqnarray}
&& \langle T(P^{\prime\prime},\varepsilon^{\prime\prime*})|A_\mu| B_{(s)}(P^\prime)\rangle
\nonumber\\&&=-i\Big\{k(q^2)\varepsilon^{\prime\prime*}_{\mu\nu}P^\nu +\varepsilon^{\prime\prime*}_{\alpha\beta}P^\alpha P^\beta[P_\mu b_+(q^2)
  +q_\mu b_-(q^2)]\Big\},\label{eq:B TO T}
 \end{eqnarray}
where $V$, $S$, $A$, and $T$ on the left-hand side of the equations denote vector, scalar, axial-vector, and tensor charmed mesons, respectively. $V_\mu$ and $A_\mu$ are the vector current $\bar c\gamma_\mu b$ and axial-vector current $\bar c\gamma_\mu\gamma_5 b$, respectively. The conventions $P=P^\prime+P^{\prime\prime}$, $q=P^\prime-P^{\prime\prime}$, and $\epsilon_{0123}=1$ are adopted. In addition, there exist two relations among these form factors
 \begin{eqnarray}
 A_3^{B_{(s)}\to V}(q^2)&=&\frac{m_{B_{(s)}}+m_V}{ 2m_V}A_1^{B_{(s)}\to V}(q^2)\nonumber\\&&
 -\frac{m_{B_{(s)}}-m_{V}}{2m_{V}}\,A_2^{B_{(s)}\to V}(q^2),
 \end{eqnarray}
 \begin{eqnarray}
 V_3^{B_{(s)}\to A}(q^2)&=&\frac{m_{B_{(s)}}-m_{A}}{ 2m_{A}}
 V_1^{B_{(s)}\to A}(q^2) \nonumber
\\&&-\frac{m_{B_{(s)}}+m_{A}}{2m_{A}}\,V_2^{B_{(s)}\to A}(q^2),\label{eq:relation}
\end{eqnarray}
where we need to specify that these form factors are
dimensionless. All these form factors were deduced in
Refs. \cite{Jaus:1999zv,Cheng:2003sm} and we collect them in the
Appendix for  the readers' convenience.

\section{Numerical results}\label{sec3}

\subsection{Form factors}\label{sec31}

As explained in the Introduction, the experimentally observed  $D_{sJ}(3040)$, $D(3000)$, and $D^*(3000)$ are good candidates for the $2P$ states in the charmed or charmed-strange family, i.e., $D_{sJ}(3040)$ is explained as the first radial excitation of $D_{s1}(2460)$, which is the $1^+$ state in the $S$ doublet \cite{Sun:2009tg}. $D(3000)$ and $D^*(3000)$ can be seen as the first radial excitations of $D_1(2430)$ ($J^P=1^+$) and $D_0^*(2400)$ ($J^P=0^+$), respectively \cite{Sun:2013qca}. Before discussing their productions via the semileptonic decays of $B$ and $B_s$ mesons, we list the experimental values of the masses of $D_{sJ}(3040)$, $D(3000)$, and $D^*(3000)$, as well as their theoretical masses and that of their partners (see Table \ref{spectrum} for more details). Their parters have yet to be observed experimentally, and  thus we adopt the mass values estimated in Ref. \cite{Ebert:2009ua} in our calculation.

\renewcommand{\arraystretch}{1.3}
\begin{table*}[htbp]\centering
\caption{The experimental values of the masses of $D_{sJ}(3040)$, $D(3000)$, and $D^*(3000)$, and the theoretical masses of their partners (in units of MeV). To distinguishing two $2P_1$ states, we use the superscripts $1/2$ and $3/2$, corresponding to quantum numbers $s_\ell=1/2$ and $3/2$, respectively.}\label{spectrum}
\begin{tabular}{c|ccc|ccc}
\toprule[1pt]
 \multirow{2}{*}{$n^{2S+1}L_J$}
   &\multicolumn{3}{ c|}{$D$ meson}&\multicolumn{3}{ c }{$D_s$ meson}\\
   &Ref. \cite{Ebert:2009ua} &Ref. \cite{Di Pierro:2001uu} &Expt. &Ref. \cite{Ebert:2009ua}&Ref. \cite{Di Pierro:2001uu}&Expt. \\  \midrule[1pt]
$2^3P_0$&2919&2949&$3008.1\pm4.0$ \cite{Aaij:2013sza}&3054&3067&-- \\
$2P_1^{1/2}$&3021&3045&$2971.8\pm8.7$ \cite{Aaij:2013sza}&3154&3165 &$3044\pm8^{+30}_{-5}$ \cite{Aubert:2009ah} \\
$2P_1^{3/2}$&2932&2995&--&3067&3114 &--\\
$2^3P_2$&3012&3035&--&3142&3157 &-- \\
\bottomrule[1pt]
\end{tabular}
\end{table*}

Under the heavy-quark effective theory, the light degrees of freedom are decoupled from the heavy quark in a heavy-meson system,
 which implies that the angular momenta of the light degrees of freedom ($s_\ell$) and the heavy quark are conserved separately.
 With these good quantum numbers, we can categorize the heavy mesons into several doublets. For example, the $S$ doublet is
 ($0^+$,$1^+$) with $s_\ell=\frac{1}{2}$, and the $T$ doublet is ($1^+$,$2^+$) with $s_\ell=\frac{3}{2}$. Thus,
 we can label $1^+$ states in the $S$ and $T$ doublets as $P^{1/2}_1$ and $P^{3/2}_1$, respectively,
 which satisfy the following relations \cite{Close:2005se}
\begin{eqnarray}
 |P^{3/2}_1\rangle &=& \sqrt{\frac{2}{3}} |^1P_1\rangle +\sqrt{\frac{1}{3}}
 |^3P_1\rangle,\label{1}\\
 |P^{1/2}_1\rangle &=& \sqrt{\frac{1}{3}} |^1P_1\rangle -\sqrt{\frac{2}{3}}
 |^3P_1\rangle.\label{12}
\end{eqnarray}

We apply the light-front wave functions in LFWFs when calculating the form factors, which can be obtained by solving the realistic bound-state equations. Practically, we use a Gaussian-type wave function for convenience, i.e.,
\begin{eqnarray*}
 \varphi^\prime
    &=&\varphi^\prime(x_2,p^\prime_\perp)
             =4 \left({\pi\over{\beta^{\prime2}}}\right)^{3/4}
               \sqrt{{dp^\prime_z\over{dx_2}}}~{\rm exp}
               \left(-{p^{\prime2}_z+p^{\prime2}_\bot\over{2 \beta^{\prime2}}}\right),
\\
 \varphi^\prime_p
    &=&\varphi^\prime_p(x_2,p^\prime_\perp)=\sqrt{2\over{\beta^{\prime2}}}
    ~\varphi^\prime,\quad
         \frac{dp^\prime_z}{dx_2}=\frac{e^\prime_1 e_2}{x_1 x_2
         M^\prime_0}.
 \label{wf}
\end{eqnarray*}
For these $2P$ states in the $D$ and $D_s$ meson families, we adopt two different wave functions, i.e., the harmonic-oscillator form given in Refs. \cite{Li:2010bb,Wang:2010npa} and the modified harmonic-oscillator function as suggested in Ref. \cite{Ke:2010vn}, which correspond to
\begin{eqnarray}
 \varphi^\prime_p
    &=&\varphi^\prime_p(x_2,p^\prime_\perp)\nonumber\\
    &=&\sqrt{2\over{\beta^{\prime2}}}\varphi^\prime_{2S}(x_2,p^\prime_\perp)\nonumber\\
             &=&4\sqrt
 {\frac{2}{3}}\left(\frac{\pi}{\beta^{\prime2}}\right)^{3/4} \sqrt{\frac{2}{{\beta^{\prime2}}}}
 \sqrt{\frac{dp_z}{dx_2}}{\rm exp}\left(-\frac{p_{\bot}^2+p_z^2}{2\beta^{\prime2}}\right)\nonumber\\&&
 \times\left(\frac{p_{\bot}^2+p_z^2}{\beta^{\prime2}}-\frac{3}{2}\right),\label{wf2p}
\end{eqnarray}
and
\begin{eqnarray}
 \varphi^\prime_{p(M)}
    &=&\varphi^\prime_{p(M)}(x_2,p^\prime_\perp)\nonumber\\
    &=&\sqrt{2\over{\beta^{\prime2}}}\varphi^\prime_{2S(M)}(x_2,p^\prime_\perp)\nonumber\\
             &=&4\left(\frac{\pi}{\beta^{\prime2}}\right)^{3/4} \sqrt{\frac{2}{{\beta^{\prime2}}}}
 \sqrt{\frac{dp_z}{dx_2}}{\rm exp}\left(-\frac{2^{1/1.82}}{2}\frac{p_{\bot}^2+p_z^2}{\beta^{\prime2}}\right)\nonumber\\&&
 \times\left(1.55\frac{p_{\bot}^2+p_z^2}{\beta^{\prime2}}-1.89\right),\label{wf2pM}
\end{eqnarray}
respectively.


In our calculation, the constituent quark masses are taken as $m_{u,d}=0.26$ GeV, $m_s=0.37$ GeV, $m_c=1.40$ GeV and $m_b=4.64$ GeV
\cite{Cheng:2003sm}. In addition, the shape parameter $\beta$ in LFWFs can be determined by the corresponding decay constants \cite{Cheng:2003sm}. We adopt the lattice results for the $B$ meson and $B_s$ meson, $f_B=190$ MeV, and $f_{B_s}=231$ MeV \cite{Gamiz:2009ku}, which allows us to estimate the corresponding shape parameters, i.e., $\beta_B=0.549$ GeV and  $\beta_{B_s}=0.6224 $ GeV. For the $P$-wave $D_{(s)}$ mesons, {we adopt
$\beta_{D^*(3000)}=\beta_{D(3000)}=\beta_{D(2P^{3/2}_1)}=\beta_{D(2^3P_2)}=0.27\pm0.03$ GeV and  $\beta_{D_s(2^3P_0)}=\beta_{D_s(3040)}=\beta_{D_s(2P^{3/2}_1)}=\beta_{D_s(2^3P_2)}=0.3\pm0.03$ GeV corresponding to the wave function listed in Eq. (\ref{wf2p}), while $\beta_{D^*(3000)}=\beta_{D(3000)}=\beta_{D(2P^{3/2}_1)}=\beta_{D(2^3P_2)}=0.39\pm0.03$ GeV and  $\beta_{D_s(2^3P_0)}=\beta_{D_s(3040)}=\beta_{D_s(2P^{3/2}_1)}=\beta_{D_s(2^3P_2)}=0.41\pm0.03$ GeV corresponding to the modified wave function shown in Eq. (\ref{wf2pM}). With these $\beta$ values, we can reproduce the obtained decay constants given in Ref. \cite{Wang:2007av}.}

Since we take $q^+=0$ in the light-front quark model, all obtained results for the form factors are only valid for the $q^2\le0$ region which is due to the fact that $q^2=q^+q^--q_\bot^2$. Thus, we cannot directly apply the obtained form factors to calculations of the decay width. Considering the fact that the semileptonic decay exists in the timelike region, we need to extrapolate our result to the timelike region, where we use the parametrized formula
\begin{eqnarray}
 F(q^2)=\frac{F(0)}{1-aq^2/m_{B_{(s)}}^2+b(q^2/m_{B_{(s)}}^2)^2}
\end{eqnarray}
for the timelike region. $F(q^2)$ stands for a form factor, while $a$ and $b$ are fixed by fitting the corresponding results for the form factor in the spacelike region ($-20\, {\rm GeV}<q^2<0\, {\rm{GeV}}$).
Finally, the obtained results are collected in Table \ref{FF}. Our obtained results for the form factor of the $B_s\to D_{sJ}(3040)$ and $B_s\to D_s(2P^{3/2}_1)$ transition matrix elements using harmonic-oscillator light-front wave functions are consistent with the results in Ref. \cite{Li:2010bb}.

{Our results show that the form factor of the semileptonic decays of the $B$ meson into the $2P$ state in the charmed meson family
is similar to that of the $B_s$ meson into the $2P$ state in the charmed-strange meson family, which reflects the SU(3) flavor symmetry. In case I, only $V_0^{B\to D(3000)}$, $V_0^{B_s\to D_{sJ}(3040)}$ and $k^{B_{(s)}\to D_{(s)}(2^3P_2)}$ are sensitive to $\beta$, while other form factors are not sensitive to the value of $\beta$.
In case II, all the form factors are not sensitive to $\beta$. This phenomenon is due to the choice of the wave function, where it is obvious that we suggest adopting the modified harmonic-oscillator wave function. In the following, we still present the results of the decay width in both cases, which will further show that different choices for the wave function can result in different situations regarding the sensitivity of the decay width to $\beta$.}

\renewcommand{\arraystretch}{1.5}
\begin{table*}[htbp]
\caption{The form factors for the semileptonic decays of $B_{(s)}$ into the corresponding $2P$ states of $D_{(s)}$ meson families. Here, $D^*(3000)$, $D(3000)$ and $D_{sJ}(3040)$ decay as $D(2^3P_0)$, $D(2P_1^{1/2})$ and  $D_s(2P_1^{1/2})$ \cite{Sun:2013qca,Sun:2009tg}, respectively.  Cases I  and II correspond to the results using the harmonic-oscillator light-front wave function in Eq. (\ref{wf2p}) and the modified harmonic-oscillator light-front wave function in Eq. (\ref{wf2pM}), respectively.}
\label{FF}
\begin{center}
\begin{tabular}{ccccc|ccccc} \toprule[1pt]
\multicolumn{10}{c}{Case I}\\
\midrule[1pt]
    & $F(q^2=0)$ & $F(q^2_{max})$ & $a$ & $b$ &  & $F(q^2=0)$ & $F(q^2_{max})$ & $a$ & $b$ \\ \midrule[1pt]
  $F_0^{B\to D^*(3000)}$ & $0.37^{+0.04}_{-0.05}$ & $0.34^{+0.03}_{-0.05}$ & $-0.42^{+0.0}_{-0.16}$ & $0.295^{+0.22}_{-0.0}$ & $F_0^{B_s\to D_s(2^3P_0)}$ &$0.41^{+0.04}_{-0.05}$ &$0.37^{+0.03}_{-0.04}$ &$-0.47^{+0.0}_{-0.13}$ &$0.37^{+0.19}_{-0.0}$ \\
  $F_1^{B\to D^*(3000)}$ & $0.37^{+0.04}_{-0.05}$ & $0.47^{+0.08}_{-0.06}$ & $1.2^{+0.2}_{-0.0}$ & $0.34^{+0.0}_{-0.04}$ & $F_1^{B_s\to D_s(2^3P_0)}$ &$0.41^{+0.04}_{-0.05}$ &$0.54^{+0.06}_{-0.07}$ &$1.35^{+0.04}_{-0.05}$ &$0.34^{+0.01}_{-0.03}$ \\
  $A^{B\to D(3000)}$ & $-0.15^{+0.02}_{-0.03}$ & $-0.20^{+0.05}_{-0.04}$ & $1.2^{+0.2}_{-0.18}$ & $0.11^{+0.07}_{-0.0}$ & $A^{B_s\to D_{sJ}(3040)}$ &$-0.17^{+0.02}_{-0.02}$ &$-0.22^{+0.04}_{-0.03}$ &$1.27^{+0.1}_{-0.17}$ &$0.16^{+0.04}_{-0.01}$ \\
  $V_0^{B\to D(3000)}$ & $0.063^{+0.032}_{-0.035}$ & $0.099^{+0.04}_{-0.071}$ & $2.2^{+0.0}_{-0.6}$ & $1.5^{+7.4}_{-0.5}$ & $V_0^{B_s\to D_{sJ}(3040)}$ &$0.08^{+0.03}_{-0.033}$ &$0.12^{+0.05}_{-0.062}$ &$2.11^{+0.0}_{-0.2}$ &$1.85^{+3.42}_{-0.91}$ \\
  $V_1^{B\to D(3000)}$ & $-0.40^{+0.02}_{-0.02}$ & $-0.35^{+0.03}_{-0.01}$ & $-0.72^{+0.01}_{-0.06}$ & $0.53^{+0.09}_{-0.01}$ & $V_1^{B_s\to D_{sJ}(3040)}$ &$-0.43^{+0.01}_{-0.01}$ &$-0.36^{+0.01}_{-0.0}$ &$-0.90^{+0.06}_{-0.08}$ &$0.73^{+0.06}_{-0.07}$ \\
  $V_2^{B\to D(3000)}$ & $-0.16^{+0.03}_{-0.03}$ & $-0.2^{+0.05}_{-0.05}$ & $1.2^{+0.18}_{-0.28}$ & $0.2^{+0.02}_{-0.0}$ & $V_2^{B_s\to D_{sJ}(3040)}$ &$-0.18^{+0.03}_{-0.02}$ &$-0.23^{+0.05}_{-0.04}$ &$1.22^{+0.14}_{-0.2}$ &$0.19^{+0.03}_{-0.01}$ \\
  $A^{B
  \to D(2P^{3/2}_1)}$ & $0.27^{+0.01}_{-0.03}$ & $0.35^{+0.03}_{-0.03}$ & $1.35^{+0.07}_{-0.03}$ & $0.5^{+0.0}_{-0.14}$ & $A^{B_s\to D_s(2P^{3/2}_1)}$ &$0.26^{+0.02}_{-0.02}$ &$0.35^{+0.02}_{-0.03}$ &$1.41^{+0.01}_{-0.05}$ &$0.41^{+0.04}_{-0.04}$ \\
  $V_0^{B\to D(2P^{3/2}_1)}$ & $0.56^{+0.04}_{-0.05}$ & $0.81^{+0.06}_{-0.09}$ & $1.7^{+0.0}_{-0.12}$ & $0.6^{+0.03}_{-0.12}$ & $V_0^{B_s\to D_s(2P^{3/2}_1)}$ &$0.57^{+0.04}_{-0.05}$ &$0.80^{+0.05}_{-0.08}$ &$1.65^{+0.0}_{-0.05}$ &$0.47^{+0.07}_{-0.06}$ \\
  $V_1^{B\to D(2P^{3/2}_1)}$ & $0.98^{+0.02}_{-0.04}$ & $0.97^{+0.02}_{-0.03}$ & $-0.027^{+0.001}_{-0.025}$ & $0.22^{+0.0}_{-0.06}$ & $V_1^{B_s\to D_s(2P^{3/2}_1)}$ &$1.08^{+0.02}_{-0.05}$ &$1.07^{+0.03}_{-0.03}$ &$0.01^{+0.04}_{-0.056}$ &$0.18^{+0.02}_{-0.01}$ \\
  $V_2^{B\to D(2P^{3/2}_1)}$ & $-0.12^{+0.024}_{-0.02}$ & $-0.086^{+0.016}_{-0.054}$ & $1.1^{+1.0}_{-0.0}$ & $15.1^{+2.7}_{-4.6}$ & $V_2^{B_s\to D_s(2P^{3/2}_1)}$ &$-0.12^{+0.02}_{-0.02}$ &$-0.12^{+0.04}_{-0.03}$ &$2.0^{+0.14}_{-0.2}$ &$12.3^{+3.6}_{-2.6}$ \\
  $h^{B\to D(2^3P_2)}$ & $0.022^{+0.003}_{-0.003}$ & $0.032^{+0.004}_{-0.005}$ & $1.87^{+0.03}_{-0.07}$ & $1.22^{+0.03}_{-0.03}$ & $h^{B_s\to D_s(2^3P_2)}$ &$0.023^{+0.002}_{-0.003}$ &$0.032^{+0.003}_{-0.004}$ &$1.88^{+0.01}_{-0.02}$  &$1.2^{+0.02}_{-0.02}$ \\
  $k^{B\to D(2^3P_2)}$ & $0.92^{+0.32}_{-0.41}$ & $1.06^{+0.47}_{-0.56}$ & $0.84^{+0.3}_{-0.77}$ & $0.65^{+0.34}_{-0.09}$ & $k^{B_s\to D_s(2^3P_2)}$ &$0.95^{+0.28}_{-0.34}$ &$1.13^{+0.4}_{-0.47}$ &$1.03^{+0.19}_{-0.44}$ &$0.67^{+0.16}_{-0.09}$ \\
  $b_+^{B\to D(2^3P_2)}$ & $-0.017^{+0.001}_{-0.001}$ & $-0.026^{+0.002}_{-0.0}$ & $1.83^{+0.01}_{-0.04}$ & $1.04^{+0.0}_{-0.08}$ & $b_+^{B_s\to D_s(2^3P_2)}$ &$-0.018^{+0.002}_{-0.0}$ &$-0.025^{+0.002}_{-0.0}$ &$1.8^{+0.03}_{-0.02}$ &$1.05^{+0.04}_{-0.12}$ \\
  $b_-^{B\to D(2^3P_2)}$ & $0.024^{+0.004}_{-0.005}$ & $0.033^{+0.005}_{-0.008}$ & $1.65^{+0.1}_{-0.22}$ & $1.21^{+0.03}_{-0.05}$ & $b_-^{B_s\to D_s(2^3P_2)}$ &$0.024^{+0.003}_{-0.004}$ &$0.032^{+0.005}_{-0.006}$ &$1.71^{+0.03}_{-0.13}$ &$1.31^{+0.03}_{-0.03}$ \\ \bottomrule[1pt]
\multicolumn{10}{c}{Case II}\\
\midrule[1pt]
    & $F(q^2=0)$ & $F(q^2_{max})$ & $a$ & $b$ &  & $F(q^2=0)$ & $F(q^2_{max})$ & $a$ & $b$ \\ \midrule[1pt]
  $F_0^{B\to D^*(3000)}$ & $0.31^{+0.01}_{-0.02}$ & $0.27^{+0.01}_{-0.01}$ & $-0.69^{+0.05}_{-0.07}$ & $0.64^{+0.05}_{-0.05}$ & $F_0^{B_s\to D_s(2^3P_0)}$ &$0.33^{+0.01}_{-0.02}$ &$0.28^{+0.01}_{-0.01}$ &$-0.68^{+0.05}_{-0.05}$ &$0.63^{+0.05}_{-0.04}$ \\
  $F_1^{B\to D^*(3000)}$ & $0.31^{+0.01}_{-0.02}$ & $0.42^{+0.01}_{-0.02}$ & $1.47^{+0.0}_{-0.04}$ & $0.32^{+0.01}_{-0.02}$ & $F_1^{B_s\to D_s(2^3P_0)}$ &$0.32^{+0.02}_{-0.01}$ &$0.44^{+0.02}_{-0.03}$ &$1.45^{+0.0}_{-0.02}$ &$0.30^{+0.02}_{-0.02}$ \\
  $A^{B\to D(3000)}$ & $-0.14^{+0.01}_{-0.01}$ & $-0.20^{+0.02}_{-0.01}$ & $1.56^{+0.0}_{-0.04}$ & $0.26^{+0.01}_{-0.01}$ & $A^{B_s\to D_{sJ}(3040)}$ &$-0.14^{+0.01}_{-0.01}$ &$-0.19^{+0.01}_{-0.02}$ &$1.49^{+0.02}_{-0.05}$ &$0.16^{+0.0}_{-0.0}$ \\
  $V_0^{B\to D(3000)}$ & $0.102^{+0.018}_{-0.017}$ & $0.148^{+0.016}_{-0.02}$ & $1.75^{+0.15}_{-0.18}$ & $0.68^{+0.35}_{-0.22}$ & $V_0^{B_s\to D_{sJ}(3040)}$ &$0.107^{+0.019}_{-0.0}$ &$0.156^{+0.019}_{-0.025}$ &$1.84^{+0.15}_{-0.17}$ &$0.86^{+0.49}_{-0.29}$ \\
  $V_1^{B\to D(3000)}$ & $-0.27^{+0.0}_{-0.01}$ & $-0.23^{+0.01}_{-0.0}$ & $-0.92^{+0.04}_{-0.05}$ & $0.75^{+0.04}_{-0.05}$ & $V_1^{B_s\to D_{sJ}(3040)}$ &$-0.27^{+0.0}_{-0.0}$ &$-0.22^{+0.01}_{-0.0}$ &$-1.15^{+0.05}_{-0.07}$ &$0.94^{+0.06}_{-0.05}$ \\
  $V_2^{B\to D(3000)}$ & $-0.15^{+0.01}_{-0.01}$ & $-0.21^{+0.02}_{-0.02}$ & $1.56^{+0.03}_{-0.07}$ & $0.29^{+0.02}_{-0.02}$ & $V_2^{B_s\to D_{sJ}(3040)}$ &$-0.15^{+0.01}_{-0.01}$ &$-0.21^{+0.02}_{-0.02}$ &$1.49^{+0.04}_{-0.08}$ &$0.20^{+0.01}_{-0.0}$ \\
  $A^{B
  \to D(2P^{3/2}_1)}$ & $0.20^{+0.0}_{-0.0}$ & $0.28^{+0.0}_{-0.01}$ & $1.46^{+0.01}_{-0.04}$ & $0.34^{+0.03}_{-0.03}$ & $A^{B_s\to D_s(2P^{3/2}_1)}$ &$0.20^{+0.0}_{-0.01}$ &$0.27^{+0.0}_{-0.01}$ &$1.46^{+0.0}_{-0.03}$ &$0.34^{+0.03}_{-0.04}$ \\
  $V_0^{B\to D(2P^{3/2}_1)}$ & $0.42^{+0.01}_{-0.01}$ & $0.63^{+0.0}_{-0.01}$ & $1.78^{+0.06}_{-0.08}$ & $0.52^{+0.1}_{-0.1}$ & $V_0^{B_s\to D_s(2P^{3/2}_1)}$ &$0.42^{+0.01}_{-0.02}$ &$0.61^{+0.01}_{-0.02}$ &$1.76^{+0.04}_{-0.07}$ &$0.42^{+0.07}_{-0.08}$ \\
  $V_1^{B\to D(2P^{3/2}_1)}$ & $0.66^{+0.01}_{-0.01}$ & $0.64^{+0.01}_{-0.02}$ & $-0.13^{+0.064}_{-0.08}$ & $0.22^{+0.04}_{-0.03}$ & $V_1^{B_s\to D_s(2P^{3/2}_1)}$ &$0.73^{+0.0}_{-0.0}$ &$0.71^{+0.01}_{-0.01}$ &$-0.11^{+0.06}_{-0.07}$ &$0.22^{+0.03}_{-0.03}$ \\
  $V_2^{B\to D(2P^{3/2}_1)}$ & $-0.11^{+0.01}_{-0.01}$ & $-0.13^{+0.02}_{-0.02}$ & $2.19^{+0.0}_{-0.03}$ & $7.9^{+1.8}_{-1.3}$ & $V_2^{B_s\to D_s(2P^{3/2}_1)}$ &$-0.11^{+0.01}_{-0.01}$ &$-0.12^{+0.02}_{-0.02}$ &$2.21^{+0.0}_{-0.04}$ &$8.2^{+1.8}_{-1.4}$ \\
  $h^{B\to D(2^3P_2)}$ & $0.018^{+0.001}_{-0.001}$ & $0.027^{+0.001}_{-0.002}$ & $1.93^{+0.01}_{-0.04}$ & $1.17^{+0.05}_{-0.06}$ & $h^{B_s\to D_s(2^3P_2)}$ &$0.018^{+0.001}_{-0.001}$ &$0.026^{+0.001}_{-0.002}$ &$1.92^{+0.01}_{-0.03}$  &$1.16^{+0.05}_{-0.06}$ \\
  $k^{B\to D(2^3P_2)}$ & $1.32^{+0.11}_{-0.14}$ & $1.7^{+0.16}_{-0.21}$ & $1.32^{+0.04}_{-0.09}$ & $0.61^{+0.05}_{-0.06}$ & $k^{B_s\to D_s(2^3P_2)}$ &$1.19^{+0.12}_{-0.15}$ &$1.51^{+0.17}_{-0.21}$ &$1.35^{+0.04}_{-0.08}$ &$0.67^{+0.08}_{-0.07} $\\
  $b_+^{B\to D(2^3P_2)}$ & $-0.013^{+0.001}_{-0.0}$ & $-0.018^{+0.001}_{-0.0}$ & $1.77^{+0.07}_{-0.1}$ & $0.87^{+0.1}_{-0.1}$ & $b_+^{B_s\to D_s(2^3P_2)}$ &$-0.013^{+0.0}_{-0.0}$ &$-0.018^{+0.001}_{-0.0}$ &$1.78^{+0.06}_{-0.09}$ &$0.87^{+0.09}_{-0.1}$ \\
  $b_-^{B\to D(2^3P_2)}$ & $0.023^{+0.001}_{-0.001}$ & $0.033^{+0.001}_{-0.003}$ & $1.82^{+0.01}_{-0.04}$ & $1.27^{+0.03}_{-0.04}$ & $b_-^{B_s\to D_s(2^3P_2)}$ &$0.022^{+0.001}_{-0.002}$ &$0.029^{+0.003}_{-0.001}$ &$1.76^{+0.08}_{-0.0}$ &$1.40^{+0.0}_{-0.13}$ \\ \bottomrule[1pt]
\end{tabular}
\end{center}
\end{table*}

\subsection{The semileptonic decay widths}

Using these obtained form factors, we can calculate the decay widths of the production of these $2P$ states in the $D_{(s)}$ families via the $B_{(s)}$ semileptonic decays. The concrete expressions for  these semileptonic decays can be obtained by using the helicity amplitude, i.e., the decay width of the scalar $D_{(s)}$ is
\begin{eqnarray}
&& \frac{d\Gamma(B_{(s)}\to S\ell\bar\nu)}{dq^2}\nonumber\\&& =\left(\frac{q^2-m_\ell^2}{q^2}\right)^2\frac{ {\sqrt{\lambda(m_{B_{(s)}}^2,m_S^2,q^2)}} G_F^2 V_{cb}^2} {384m_{B_{(s)}}^3\pi^3}\frac{1}{q^2} \nonumber\\&&
\quad \times \Big\{ (m_\ell^2+2q^2) \lambda(m_{B_{(s)}}^2,m_S^2,q^2) [F_1^{B{(s)}\to S}(q^2)]^2 \nonumber\\&&
\quad +3 m_\ell^2(m_{B_{(s)}}^2-m_S^2)^2 [F_0^{B{(s)}\to S}(q^2)]^2\Big\},
\end{eqnarray}
and the decay width of the axial-vector $D_{(s)}$ is
\begin{eqnarray}
&&  \frac{d\Gamma(B_{(s)}\to A\ell\bar\nu)}{dq^2}\nonumber\\&&= \frac{d\Gamma_L(B_{(s)}\to A\ell\bar\nu)}{dq^2}+
 \frac{d\Gamma^+(B_{(s)}\to A\ell\bar\nu)}{dq^2}\nonumber\\&&
\quad +\frac{d\Gamma^-(B_{(s)}\to A\ell\bar\nu)}{dq^2}
\end{eqnarray},
with
\begin{eqnarray}
&& \frac{d\Gamma_L(B_{(s)}\to A\ell\bar\nu)}{dq^2}\nonumber\\&&=\left(\frac{q^2-m_\ell^2}{q^2}\right)^2\frac{ {\sqrt{\lambda(m_{B_{(s)}}^2,m_A^2,q^2)}}
 G_F^2 V_{cb}^2} {384m_{B_{(s)}}^3\pi^3}\frac{1}{q^2} \nonumber\\&&
 \quad\times \Bigg\{ 3 m_\ell^2 \lambda(m_{B_{(s)}}^2,m_A^2,q^2) [V_0^{B_{(s)}\to A}(q^2)]^2\nonumber\\
 && \quad+ (m_\ell^2+2q^2) \Bigg|\frac{1}{2m_A}  \Big[
 (m_{B_{(s)}}^2-m_A^2-q^2) \nonumber\\&&\quad
 \times(m_{B_{(s)}}-m_A)V_1^{B_{(s)}\to A}(q^2)-\frac{\lambda(m_{B_{(s)}}^2,m_A^2,q^2)}{m_{B_{(s)}}-m_A} \nonumber\\&&
\quad \times V_2^{B_{(s)}\to A}(q^2)\Big]\Bigg|^2
 \Bigg\},
 \end{eqnarray}
\begin{eqnarray}
 &&\frac{d\Gamma^\pm(B_{(s)}\to A\ell\bar\nu)}{dq^2}\nonumber\\&&=\left(\frac{q^2-m_\ell^2}{q^2}\right)^2\frac{
 {\sqrt{\lambda(m_{B_{(s)}}^2,m_A^2,q^2)}} G_F^2V_{cb}^2}{384m_{B_{(s)}}^3\pi^3}
 \nonumber\\
 && \quad\times \Bigg\{ (m_\ell^2+2q^2) \lambda(m_{B_{(s)}}^2,m_A^2,q^2)\Bigg|\frac{A^{B_{(s)}\to A}(q^2)}{m_{B_{(s)}}-m_A} \nonumber\\&&\quad
 \mp\frac{(m_{B_{(s)}}-m_A)V_1^{B_{(s)}\to A}(q^2)}{\sqrt{\lambda(m_{B_{(s)}}^2,m_A^2,q^2)}}\Bigg|^2
 \Bigg\},
\end{eqnarray}
where $\pm$ and $L$ are the polarizations of the axial-vector $D_{(s)}$ meson, and $m_\ell$ is the mass of the lepton. We define $\lambda(a^2,b^2,c^2)=(a^2-b^2-c^2)^2-4b^2c^2$.

In Ref. \cite{Wang:2009mi},  a special way of calculating the semileptonic decay width of $B_{(s)}$ into the tensor $D_{(s)}$ meson was proposed. With the new definition of the form factors listed in Eq. (\ref{B TO T}), we can easily obtain the corresponding decay width \cite{Wang:2009mi}:
\begin{eqnarray}
&&\frac{d\Gamma(B_{(s)}\to T\ell\bar\nu)}{dq^2}\nonumber\\&&= \frac{d\Gamma_L(B_{(s)}\to Tl\ell\bar\nu)}{dq^2}+
 \frac{d\Gamma^+(B_{(s)}\to T\ell\bar\nu)}{dq^2}\nonumber\\&&
 \quad+\frac{d\Gamma^-(B_{(s)}\to T\ell\bar\nu)}{dq^2}
 \end{eqnarray}
with
\begin{eqnarray}
 && \frac{d\Gamma_L(B_{(s)}\to T\ell\bar\nu)}{dq^2}\nonumber\\&&=\frac{1}{2}\frac{\lambda(m_{B_{(s)}}^2,m_T^2,q^2)}{4m_T^2} \frac{d\Gamma_L(B_{(s)}\to
 A\ell\bar\nu)}{dq^2}\Bigg|_{V_{0,1,2}^{B_{(s)}\to A}\to V_{0,1,2}^{B_{(s)}\to T}},
 \end{eqnarray}
and
\begin{eqnarray}
&& d\Gamma^{\pm}(B_{(s)}\to T\ell\bar\nu)\nonumber\\&&=\frac{2}{3}\frac{\lambda(m_{B_{(s)}}^2,m_T^2,q^2)}{4m_T^2}\nonumber\\&&
 \quad\times \frac{d\Gamma^{\pm}(B_{(s)}\to
 A\ell\bar\nu)}{dq^2}\Bigg|_{(V_{1}^{B_{(s)}\to A},A^{B_{(s)}\to A})\to (V_{1}^{B_{(s)}\to T},A^{B_{(s)}\to T})}.
\end{eqnarray}

With the above preparation, we can calculate the branching ratios of these discussed semileptonic decays, which are collected in Table \ref{decay}.  Here, the obtained breaching ratios for $B_s\to D_{sJ}(3040)\ell \nu_\ell$ and $B_s\to D_s(2P^{3/2}_1)\ell \nu_\ell$ using the harmonic-oscillator light-front wave functions are consistent with the results given in Ref. \cite{Li:2010bb}.
{In case I, the branching ratios of $B_{(s)}\to D_{(s)}(2^3P_2)\ell \bar\nu_\ell$ are sensitive to $\beta$, which is different from the situation involving the branching ratios of other semileptonic decays. This fact is  consistent with the behaviors of the form factors that are dependent on $\beta$, which was discussed in Sec. \ref{sec31}. Although the form factors $V_0^{B\to D(3000)}$ and $V_0^{B_s\to D_{sJ}(3040)}$ are sensitive to the $\beta$ parameters, the branching ratios of the processes $B_{(s)} \to D_{(s)}(2P^{1/2}_1)\ell\bar\nu_\ell$ have small uncertainties. In case II, the obtained branching ratios are not sensitive to $\beta$.}

In general, the decays $B_{(s)} \to D_{(s)}e(\mu)\bar\nu_{e(\mu)}$ are 2 orders of magnitude larger than those of $B_{(s)} \to D_{(s)}\tau\bar\nu_\tau$. In addition, we also notice that the decay widths of $B_{(s)} \to D_{(s)}(2P^{3/2}_1)\ell\bar\nu_\ell$ are 10 times larger than the corresponding decay widths of $B_{(s)} \to D_{(s)}(2P^{1/2}_1)\ell\bar\nu_\ell$, which is due to the mixing angle describing the mixture between $2^3P_1$ and $2^3P_1$ states in the heavy-quark limit. From Table \ref{decay}, we see that  the semileptonic decays of $B_{(s)}$ into the corresponding $2P$ states of the $D_{(s)}$ meson families have large branching ratios, which implies that these semileptonic decays can be accessible at future experiments. As shown in Table \ref{decay}, the results for case I are similar to the corresponding results for case II, which indicates that taking two difference forms for the wave function cannot give obviously different results.

\renewcommand{\arraystretch}{1.5}
\begin{table}[htbp]
\caption{The branching ratios of the semileptonic decays of $B_{(s)}$ into the corresponding $2P$ states of $D_{(s)}$ meson families. Here, cases I  and II correspond to the results using the harmonic-oscillator light-front wave function in Eq. (\ref{wf2p}) and the modified harmonic-oscillator light-front wave function in Eq. (\ref{wf2pM}), respectively. All results are multiplied by a factor of $10^{-3}$.}
\label{decay}
\begin{center}
\begin{tabular}{lccc} \toprule[1pt]
\multicolumn{4}{c}{Case I}\\
\midrule[1pt]
  & $\ell=e$ & $\ell=\mu$ & $\ell=\tau$ \\ \midrule[1pt]
  $BR(B \to D^*(3000)\ell\bar\nu_\ell)$ & $1.37^{+0.38}_{-0.35}$ & $1.36^{+0.37}_{-0.35}$ & $0.027^{+0.006}_{-0.007}$ \\
  $BR(B_s \to D_s(2^3P_0)\ell\bar\nu_\ell)$ & $1.7^{+0.37}_{-0.36}$ & $1.7^{+0.35}_{-0.38}$ & $0.041^{+0.007}_{-0.009}$ \\
  $BR(B \to D(3000)\ell\bar\nu_\ell)$ & $0.313^{+0.095}_{-0.077}$ & $0.31^{+0.094}_{-0.076}$ & $0.0086^{+0.0014}_{-0.0015}$  \\
  $BR(B_s \to D_s(3040)\ell\bar\nu_\ell)$ & $0.359^{+0.083}_{-0.072}$ & $0.355^{+0.083}_{-0.07}$ & $0.01^{+0.001}_{-0.0012}$  \\
  $BR(B \to D(2P^{3/2}_1)\ell\bar\nu_\ell)$ & $5.01^{+0.66}_{-0.64}$ & $4.96^{+0.65}_{-0.64}$ & $0.12^{+0.01}_{-0.01}$  \\
  $BR(B_s \to D_s(2P^{3/2}_1)\ell\bar\nu_\ell)$ & $4.66^{+0.48}_{-0.61}$ & $4.61^{+0.47}_{-0.6}$ & $0.089^{+0.006}_{-0.01}$  \\
  $BR(B \to D(2^3P_2)\ell\bar\nu_\ell)$ & $0.96^{+1.02}_{-0.73}$ & $0.946^{+1.0}_{-0.721}$ & $0.0076^{+0.008}_{-0.0058}$ \\
  $BR(B_s \to D_s(2^3P_2)\ell\bar\nu_\ell)$ & $0.819^{+0.67}_{-0.537}$ & $0.807^{+0.66}_{-0.529}$ & $0.0049^{+0.0039}_{-0.0032}$ \\ \bottomrule[1pt]
\multicolumn{4}{c}{Case II}\\
\midrule[1pt]
  & $\ell=e$ & $\ell=\mu$ & $\ell=\tau$ \\ \midrule[1pt]
  $BR(B \to D^*(3000)\ell\bar\nu_\ell)$ & $1.01^{+0.07}_{-0.11}$ & $0.995^{+0.07}_{-0.108}$ & $0.0186^{+0.001}_{-0.0018}$ \\
  $BR(B_s \to D_s(2^3P_0)\ell\bar\nu_\ell)$ & $1.10^{+0.1}_{-0.12}$ & $1.09^{+0.09}_{-0.12}$ & $0.0252^{+0.0018}_{-0.002}$ \\
  $BR(B \to D(3000)\ell\bar\nu_\ell)$ & $0.257^{+0.039}_{-0.044}$ & $0.254^{+0.038}_{-0.044}$ & $0.0052^{+0.0004}_{-0.0005}$  \\
  $BR(B_s \to D_s(3040)\ell\bar\nu_\ell)$ & $0.249^{+0.04}_{-0.04}$ & $0.246^{+0.04}_{-0.042}$ & $0.0052^{+0.0004}_{-0.0005}$  \\
  $BR(B \to D(2P^{3/2}_1)\ell\bar\nu_\ell)$ & $2.72^{+0.02}_{-0.11}$ & $2.69^{+0.02}_{-0.11}$ & $0.0603^{+0.0}_{-0.002}$  \\
  $BR(B_s \to D_s(2P^{3/2}_1)\ell\bar\nu_\ell)$ & $2.42^{+0.07}_{-0.14}$ & $2.39^{+0.07}_{-0.13}$ & $0.043^{+0.0}_{-0.001}$  \\
  $BR(B \to D(2^3P_2)\ell\bar\nu_\ell)$ & $2.52^{+0.49}_{-0.57}$ & $2.48^{+0.48}_{-0.56}$ & $0.0192^{+0.0038}_{-0.0045}$ \\
  $BR(B_s \to D_s(2^3P_2)\ell\bar\nu_\ell)$ & $1.5^{+0.36}_{-0.4}$ & $1.48^{+0.36}_{-0.38}$ & $0.0086^{+0.002}_{-0.0022}$ \\ \bottomrule[1pt]
\end{tabular}
\end{center}
\end{table}

\subsection{The comparison of the present result and that in the heavy-quark limit}

\subsubsection{The relations in the heavy-quark limit}

For the processes discussed in the present work, the corresponding decay amplitudes can be expressed by the form factors. In the heavy-quark limit, more constraints exist for these processes, which further gives the relations among the obtained form factors.
In heavy-quark effective theory, the transition amplitudes of $B_{(s)}$ decays into charmed/charmed-strange mesons can be expressed by the Isgur-Wise functions. 
The former transition matrix elements given in Eqs. (\ref{eq:B TO_V1})-(\ref{eq:B TO T}) in the heavy-quark limit are equivalent to the corresponding transition matrix elements expressed by the Isgur-Wise functions. Thus, the relations between the previous form factors and the Isgur-Wise functions can be obtained, which was discussed in Refs. \cite{Cheng:2003sm,Isgur:1990jf}. Finally, there exist model-independent relations,
\begin{eqnarray}
&&\tau_{1/2}^u=\tau_{1/2}^\ell=\tau_{1/2}^q=\tau_{1/2}^c,\label{HQSR1}\\
&&\tau_{3/2}^\ell=\tau_{3/2}^{c+}=\tau_{3/2}^{c-}=\tau_{3/2}^{h}=\tau_{3/2}^{k}=\tau_{3/2}^{q}=\tau_{3/2}^{b},\label{HQSR2}\\
&&b_+(q^2)+b_-(q^2)=0,  \label{HQSR31}\\
&&c^{1/2}_+(q^2)+c^{1/2}_-(q^2)=0,  \label{HQSR32}\\
&&(m_B+m_{D^*_0})u_+(q^2)+(m_B-m_{D^*_0})u_-(q^2)=0,\label{HQSR33}
\end{eqnarray}
which are due to the constraint in the heavy-quark limit, where the expressions for the functions in Eqs. (\ref{HQSR1})-(\ref{HQSR2}) are listed in the Appendix [see Eqs. (\ref{f1})-(\ref{f2})].

In the following, we check whether the results in the light-front quark model can satisfy the relations listed in Eqs. (\ref{HQSR1})-(\ref{HQSR33}). We need to specify that we take the semileptonic decays of the $B$ meson into charmed mesons as an example to carry out the discussion.
In Tables \ref{IWf1} and \ref{IWf2}, we present the numerical results of $\tau_{1/2}^\mu$, $\tau_{1/2}^\ell$, $\tau_{1/2}^q$, $\tau_{1/2}^c$, $c_+^{1/2}$, $c_-^{1/2}$, $(m_{B_{(s)}}+m_{D_{(s)}})u_+$, $(m_{B_{(s)}}-m_{D_{(s)}})u_-$, $\tau_{3/2}^\ell$, $\tau_{3/2}^{c+}$, $\tau_{3/2}^{c-}$, $\tau_{3/2}^{h}$, $\tau_{3/2}^{k}$, $\tau_{3/2}^{q}$, $\tau_{3/2}^{b}$, $b_+$, and $b_-$ when taking the typical values $q^2=0$ and $q^2=q_{max}^2$, which are obtained using the light-front quark model.

\renewcommand{\arraystretch}{1.5}
\begin{table}[htbp]
\caption{The obtained results of $\tau_{1/2}^\mu$, $\tau_{1/2}^\ell$, $\tau_{1/2}^q$, $\tau_{1/2}^c$, $c_+^{1/2}$, $c_-^{1/2}$, $(m_{B_{(s)}}+m_{D_{(s)}})u_+$, and $(m_{B_{(s)}}-m_{D_{(s)}})u_-$ if using the typical  values $q^=0$ and $q^2=q_{max}^2$, which are obtained using the light-front quark model. For $\tau_{1/2}^\ell$, there is singularity when $q^2=q_{max}^2$, which denote by --.}
\label{IWf1}
\begin{center}
\begin{tabular}{c|cc|cc} \toprule[1pt]
  &\multicolumn{2}{ c|}{Case I}&\multicolumn{2}{ c }{Case II}\\
   & $q^2=0$ & $q^2=q^2_{max}$ & $q^2=0$ & $q^2=q^2_{max}$ \\ \midrule[1pt]
  $\tau_{1/2}^u$ & $0.22$ & $0.28$ & $0.23$ & $0.32$\\
  $\tau_{1/2}^\ell$ & $0.69$ & -- & $0.47$ & --\\
  $\tau_{1/2}^q$ & $0.26$ & $0.34$ & $0.24$ & $0.34$ \\
  $\tau_{1/2}^c$ & $0.26$ & $0.33$ & $0.25$ & $0.35$ \\
  $c^{1/2}_+$ & $-0.068$ & $-0.088$ & $-0.065$ & $-0.091$ \\
  $c^{1/2}_-$ & $0.061$ & $0.077$ & $0.063$ & $0.087$ \\
  $(m_{B_{(s)}}+m_{D_{(s)}})u_+$ & $-3.07$ & $-4.04$ & $-2.59$ & $-3.50$ \\
  $(m_{B_{(s)}}-m_{D_{(s)}})u_-$ & $1.177$ & $1.53$ & $1.21$ & $1.66$ \\ \bottomrule[1pt]
\end{tabular}
\end{center}
\end{table}

\renewcommand{\arraystretch}{1.5}
\begin{table}[htbp]
\caption{The calculated $\tau_{3/2}^\ell$, $\tau_{3/2}^{c+}$, $\tau_{3/2}^{c-}$, $\tau_{3/2}^{h}$, $\tau_{3/2}^{k}$, $\tau_{3/2}^{q}$, $\tau_{3/2}^{b}$, $b_+$, and $b_-$ using the light-front quark model. Here, we take $q^2=0$ and $q^2=q^2_{max}$ to present the results. For $\tau_{3/2}^\ell$, there is a singularity when $q^2=q_{max}^2$, which we denote by --.}
\label{IWf2}
\begin{center}
\begin{tabular}{l|cc|cc} \toprule[1pt]
  &\multicolumn{2}{ c|}{Case I}&\multicolumn{2}{ c }{Case II}\\
   & $q^2=0$ & $q^2=q^2_{max}$ & $q^2=0$ & $q^2=q^2_{max}$ \\ \midrule[1pt]
  $\tau_{3/2}^\ell$ & $2.14$ & $--$ & $1.44$ & $--$\\
  $\tau_{3/2}^q$ & $0.58$ & $0.85$ & $0.44$ & $0.66$ \\
  $\tau_{3/2}^{c+}$ & $0.47$ & $0.52$ & $0.41$ & $0.51$\\
  $\tau_{3/2}^{c-}$ & $0.47$ & $0.69$ & $0.32$ & $0.46$ \\
  $\tau_{3/2}^{h}$ & $0.54$ & $0.78$ & $0.45$ & $0.65$\\
  $\tau_{3/2}^{k}$ & $0.32$ & $0.40$ & $0.47$ & $0.65$ \\
  $\tau_{3/2}^b$ & $0.50$ & $0.71$ & $0.44$ & $0.62$ \\
  $b_+$ & $-0.018$ & $-0.026$ & $-0.013$ & $-0.018$ \\
  $b_-$ & $0.024$ & $0.033$ & $0.023$ & $0.033$ \\\bottomrule[1pt]
\end{tabular}
\end{center}
\end{table}

The results given in Table \ref{IWf1} correspond to the semileptonic decays of the $B$ meson into charmed mesons in the $(0^+,1^+)$ doublet. {Here, we adopt the central value of the $\beta$ parameter to present the results.}
We find that the results of $\tau_{1/2}^\mu$, $\tau_{1/2}^q$, and $\tau_{1/2}^c$ are similar to one another, which can approximately meet the requirement in Eq. (\ref{HQSR1}). However, the obtained $\tau_{1/2}^\ell$ is $2\sim 3$ times larger than the results of $\tau_{1/2}^\mu$, $\tau_{1/2}^q$, and $\tau_{1/2}^c$, which implies a violation of Eq. (\ref{HQSR1}), where $\tau_{1/2}^\ell$ is from the $B\to D(2P_1^{1/2})\ell \bar\nu_\ell$ process. In addition, the calculated $c_+^{1/2}$ and $c_-^{1/2}$ approximately satisfy Eq. (\ref{HQSR32}). The comparison between $(m_{B_{(s)}}+m_{D_{(s)}})u_+$ and $(m_{B_{(s)}}-m_{D_{(s)}})u_-$ shows that there is a large discrepancy, which violates the relation listed in Eq. (\ref{HQSR33}), where both $(m_{B_{(s)}}+m_{D_{(s)}})u_+$ and $(m_{B_{(s)}}-m_{D_{(s)}})u_-$ come from the $B\to D(2^3P_0)\ell \bar\nu_\ell$ process.

In the following, we discuss the results listed in Table \ref{IWf2}, which correspond to the semileptonic decays of the $B$ meson into charmed mesons in the $(1^+,2^+)$ doublet, {where we use the central values of the $\beta$ parameter.} The obtained $\tau_{3/2}^{c+}$, $\tau_{3/2}^{c-}$, $\tau_{3/2}^{h}$, $\tau_{3/2}^{k}$, $\tau_{3/2}^{q}$, $\tau_{3/2}^{b}$ are approximately equal to one another, while $\tau_{3/2}^{\ell}$ is much larger than the results of $\tau_{3/2}^{c+}$, $\tau_{3/2}^{c-}$, $\tau_{3/2}^{h}$, $\tau_{3/2}^{k}$, $\tau_{3/2}^{q}$, and $\tau_{3/2}^{b}$. The absolute value of $b_+$ is similar to that of $b_-$. The above comparison indicates that $\tau_{3/2}^{c+}$, $\tau_{3/2}^{c-}$, $\tau_{3/2}^{h}$, $\tau_{3/2}^{k}$, $\tau_{3/2}^{q}$, $\tau_{3/2}^{b}$, $b_+$, and $b_-$ can approximately satisfy Eqs. (\ref{HQSR2}) and (\ref{HQSR31}), whereas
$\tau_{3/2}^{\ell}$ cannot.

The above comparison reflects the existence of a discrepancy between the results obtained in the light-front quark
model and the expectations from the heavy-quark limit. We need to specify that the relations listed in Eqs. (\ref{HQSR1})-(\ref{HQSR33}) are obtained in the heavy-quark limit. However, for these  semileptonic processes the bottom- and charm-quark masses are finite since the quark masses of $\bar{c}$ and  $b$ quarks involved in our calculation are 1.4 GeV and  4.64 GeV, respectively, which cannot strictly meet the requirement of the heavy-quark limit ($m_{Q}\to +\infty$). Thus, the above discrepancy may be due to this fact.

If we perform the calculation using the heavy-quark effective theory (HQET), adopting the relations shown in Eqs. (\ref{HQSR1})-(\ref{HQSR33}) can simplify the whole calculation since we only need to calculate some universal Isgur-Wise functions and the remaining form factors can be obtained from Eqs. (\ref{HQSR1})-(\ref{HQSR33}).
However, this treatment is problematic if the $1/m_Q$ correction plays an important role. The former discussion gives a good example, i.e., we need to consider the $1/m_Q$ corrections to Eqs. (\ref{HQSR1})-(\ref{HQSR33}) [especially Eqs. (\ref{HQSR1})-(\ref{HQSR2})]\footnote{As indicated in the former comparison, $\tau_{1/2}^\ell$ and $\tau_{3/2}^\ell$ do not satisfy Eqs. (\ref{HQSR1}) and (\ref{HQSR2}), respectively. By the definitions listed in Eqs. (\ref{f3})-(\ref{f4}), $\tau_{1/2}^\ell$ and $\tau_{3/2}^\ell$ are related to the form factors $\ell^{1/2}(q^2)$ and $\ell^{3/2}(q^2)$ in Eq. (\ref{aaa}), where we adopt the superscripts $1/2$
and $3/2$ to distinguish form factors $\ell(q^2)$ for the $B\to D(2P^{1/2}_1)\ell \bar\nu_\ell$ and $B\to D(2P^{3/2}_1) \ell \bar\nu_\ell$ decays, respectively.}.

In addition, it is possible that the above discrepancy is partly due to the uncertainties of the results obtained in the light-front quark model. In Sec. \ref{wa}, we further discuss these possible sources of  the uncertainty on the numerical result.

In the following subsection, we calculate the branching ratios of these semileptonic decays in the light-front Quark model associated with HQET, and compare the results with those obtained using the light-front quark model\footnote{We would like to thank the anonymous referee for this valuable suggestion.}.

\subsubsection{The results under the light-front quark model associated with HQET}
The covariant light-front quark model within HQET was first proposed in Ref. \cite{Cheng:1997au} (before the covariant light-front quark model), and the authors of Ref. \cite{Cheng:2003sm} proved  that the covariant light-front quark model within HQET and the covariant light-front quark model are consistent. In the heavy-quark limit, the heavy-quark pair creation is suppressed, so the Z-graph contribution vanishes, which means that we can directly calculate the corresponding Isgur-Wise functions in the timelike region. Here, we only list our results for the decay width; details of the method and formula for the covariant light-front quark model within HQET can be found in Ref. \cite{Cheng:2003sm}.  
{Here, we adopt $\beta=0.51$ for the $B$ meson and $2P$ $D$ mesons, and $\beta=0.573$ for the $B_s$ meson and $2P$ $D_s$ mesons, where these $\beta$ values are determined by the decay constants of the $B$ and $B_s$ mesons \cite{Gamiz:2009ku}. Finally, the obtained branching ratios are listed in Table \ref{decayHQET}.}

In the following, we compare the results in Table \ref{decayHQET} with those shown in Table \ref{decay}. In case I, most of the branching ratios obtained in the light-front quark model with HQET are generally $2\thicksim3$ times larger or smaller than the corresponding results using the light-front quark model without considering HQET\footnote{However, 
the branching ratio of $B_{(s)} \to D_{(s)}(2^3P_2)\ell\bar\nu_\ell$ calculated in HQET is $3.6$ times larger than that obtained in the Light Front Quark model without considering HQET.}. 
In case II,  the situation of the branching ratios of $B_{(s)} \to D_{(s)}(2^3P_0)\ell\bar\nu_\ell$ with and without including HQET is similar to that in case I, except for $B\to D_{(s)}(2^3P_2)\ell \bar{\nu}_\ell$, where the branching ratios of $B\to D_{(s)}(2^3P_2)\ell \bar{\nu}_\ell$ calculated using the light-front quark model with and without HQET are close to each other.

In general, there are discrepancies in the results obtained in the light-front quark model with and without HQET, which reflects the fact that the $1/m_Q$ correction is important for the results calculated using the light-front quark model associated with HQET.

\renewcommand{\arraystretch}{1.5}
\begin{table}[htbp]
\caption{The branch ratios of the semileptonic decays of $B_{(s)}$ into the corresponding $2P$ states of $D_{(s)}$ meson families using the light-front quark model within HQET. Here, cases I and II correspond to the results using the harmonic-oscillator light-front wave function and the modified harmonic-oscillator light-front wave function, respectively. Here, $\ell$ denotes $e$ or $\mu$. All results are multiplied by a factor of $10^{-3}$.}
\label{decayHQET}
\begin{center}
\begin{tabular}{l|c|c} \toprule[1pt]
  & Case I & Case II \\ \midrule[1pt]
  $BR(B \to D^*(3000)\ell\bar\nu_\ell)$ & $0.65$ & $0.37$\\
  $BR(B_s \to D_s(2^3P_0)\ell\bar\nu_\ell)$ & $0.70$ & $0.41$ \\
  $BR(B \to D(3000)\ell\bar\nu_\ell)$ & $0.96$ & $0.56$\\
  $BR(B_s \to D_s(3040)\ell\bar\nu_\ell)$ & $0.96$ & $0.56$ \\
  $BR(B \to D(2P^{3/2}_1)\ell\bar\nu_\ell)$ & $2.36$ & $1.50$\\
  $BR(B_s \to D_s(2P^{3/2}_1)\ell\bar\nu_\ell)$ & $1.94$ & $1.24$ \\
  $BR(B \to D(2^3P_2)\ell\bar\nu_\ell)$ & $3.47$ & $2.23$ \\
  $BR(B_s \to D_s(2^3P_2)\ell\bar\nu_\ell)$ & $2.87$ & $1.86$ \\\bottomrule[1pt]
\end{tabular}
\end{center}
\end{table}

\subsection{The possible sources of the uncertainty of the result}\label{wa}

In this subsection, we discuss the possible sources of the uncertainty on the results including the form factors and branching ratios.

1) The observed $D(3000)$, $D^*(3000)$, and $D_{sJ}(3040)$ states can be seen as good candidates for the $2P$ states. Besides discussing their production via the $B_{(s)}$-meson semileptonic decays, we would also like study the production of their spin partners. However, their spin partners have yet to be observed experimentally, which means that their masses have not been
 measured. Thus, when estimating the branching ratios of these spin partners produced by the $B_{(s)}$-meson semileptonic decays, we use the theoretical predictions for the masses of these missing $2P$ states  \cite{Ebert:2009ua}, which is one of the sources of the uncertainty on the results presented in this work. 

2) In our calculation, $\beta$ (as the input parameter) determines the shapes of the corresponding spacial wave functions.
$\beta$ can be fixed by the decay constant of the meson. There are no experimental or lattice results for the decay constants and leptonic decay widths of the $2P$ $D_{(s)}$ mesons discussed in this work. In Ref. \cite{Wang:2007av}, the authors studied the radially excited $P$-wave $D_{(s)}$ mesons with the instantaneous Bethe-Salpeter method, which provided the theoretical values of the decay constants of the $2P$ charmed/charmed-strange mesons discussed here. Thus, in this work we determined $\beta$ by using the theoretical calculations from Ref. \cite{Wang:2007av}, which is also a possible source of uncertainty.

3) Two $1^+$ states in the charmed/charmed-strange meson family are a mixture between the $^1P_1$ and $^3P_1$
states. In this work, we use the mixing given by  Eqs. (\ref{1}) and (\ref{12}), which is the result in the heavy-quark limit \cite{Close:2005se}. In fact, a realistic mixing angle of the mixture between $^1P_1$ and $^3P_1$ deviates from
the result in the heavy-quark limit \cite{Sun:2009tg}, especially for the higher $1^+$ radial excitations, which is also an important source of uncertainty.


(4) Additionally, the uncertainty can come from the choice of the LFWF, which plays an important role in the light-front quark model. In this work, we adopted two types of LFWFs to present the results. The results obtained in the two cases are slightly different from each other.
Thus, more studies and discussions about the LFWF (especially for the excited states) are needed in the future.

More theoretical and experimental efforts are needed in order to reduce the uncertainties of the predicted results in this work, which is an intriguing research topic.

\section{Summary}

In the past decade, the charmed and charmed-strange meson families have became more and more abundant due to
the experimental observation of these higher charmed and charmed -strange states. Among newly observed charmed and charmed-strange states, there are two charmed states $D(3000)$ and $D^*(3000)$ and one charmed-strange state $D_{sJ}(3040)$ around 3 GeV. These observed states may be good candidates for the $2P$ states in the charmed and charmed-strange meson families \cite{Sun:2013qca,Sun:2009tg}.

At present, $D^{(*)}(3000)$ has only been reported in the inclusive processes $pp\to D^+\pi^-X$, $pp\to D^0\pi^+X$, and $pp\to D^{*+}\pi^-X$, while $D_{sJ}(3040)$ has been observed in the inclusive $e^+e^-$ interaction.
The $B_{(s)}$ semileptonic decays can provide a new approach to study these newly observed $D^{(*)}(3000)$ and $D_{sJ}(3040)$ states. In order to explore the discovery potential of $D^{(*)}(3000)$ and $D_{sJ}(3040)$ via the semileptonic decays of $B_{(s)}$,
in this work we studied the production of $D^{(*)}(3000)$, $D_{sJ}(3040)$ and their partners through the semileptonic decays of $B_{(s)}$ mesons, where the covariant light-front quark model was used in the calculation.

Our calculation indicates that the branching ratios of the $B_{(s)}$ semileptonic decays into the $2P$ states of the $D_{(s)}$ family are considerable. This information shows that experimental searches for $D^{(*)}(3000)$, $D_{sJ}(3040)$, and their partners via the $B_{(s)}$ semileptonic decays
is possible at future experiments. Thus, we suggest that the LHCb and the forthcoming Belle II experiments carry out studies of
$D^{(*)}(3000)$, $D_{sJ}(3040)$, and their partners through the $B_{(s)}$ semileptonic decays, which is an intriguing and important research topic that could further reveal the underlying properties of the $D^{(*)}(3000)$ and $D_{sJ}(3040)$ states.

\section*{Acknowledgements}
We would like to thank Professor Xue-Qian Li for his useful suggestions.
This project is supported by the National Natural Science
Foundation of China under Grants No. 11222547, No. 11175073, No. 11035006 and No. 11375128, the Ministry of Education of China
(FANEDD under Grant No. 200924, SRFDP under Grant No.
2012021111000, and NCET), the Fok Ying Tung Education Foundation
(No. 131006).

\section*{APPENDIX: SOME USEFUL FORMULAS}\label{app}
\begin{figure}[htpb]
\includegraphics[scale=0.7]{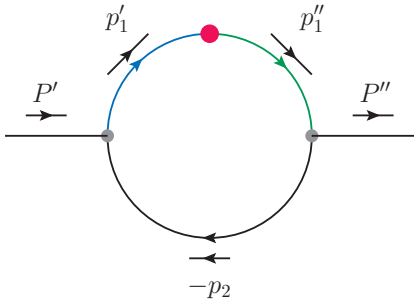}
\caption{(Color online.) The Feynman diagram for the $B_{(s)}\to
D_{(s)J}$ transition. The red point denotes the $V-A$ current
vertex, while the $b$ and $c$ quark lines are blue and
green, respectivelty.}\label{fig:feyn}
\end{figure}

The Feynman diagram for the $B_{(s)}\to D_{(s)J}$ transition is
depicted in Fig. \ref{fig:feyn}. In the calculation one needs the
light-front decomposition of the momentum, i.e.,
$P^{\prime}=(P^{\prime -}, P^{\prime +}, P^\prime_\bot)$, where
$P^{\prime\pm}=P^{\prime0}\pm P^{\prime3}$ and $P^{\prime
2}=P^{\prime +}P^{\prime -}-P^{\prime 2}_\bot$. The initial-
(final-) state meson has momentum $P^{\prime}=p_1^{\prime}+p_2$
($P^{\prime\prime}=p_1^{\prime\prime}+p_2$) and mass 
$M^\prime$ $(M^{\prime\prime})$. Here, the mass and momentum of the
antiquark inside both the initial and final mesons are $m_2$ and
$p_2$, respectively. The quark in the initial (final) meson has
 mass $m_1^{\prime(\prime\prime)}$ and momentum
$p_1^{\prime(\prime\prime)}$. These momenta are defined by the
internal variables $(x_i, p_\bot^\prime)$, i.e.,
 \begin{eqnarray}
 p_{1,2}^{\prime+}=x_{1,2} P^{\prime +},\qquad
 p^\prime_{1,2\bot}=x_{1,2} P^\prime_\bot\pm p^\prime_\bot
 \end{eqnarray}
with $x_1+x_2=1$. Taking $q^+=0$, with these variables one further
defines some useful quantities for the initial state
\begin{eqnarray}
 M^{\prime2}_0 &=&(e^\prime_1+e_2)^2=\frac{p^{\prime2}_\bot+m_1^{\prime2}} {x_1}+\frac{p^{\prime2}_{\bot}+m_2^2}{x_2}, \nonumber\\
 \widetilde M^\prime_0 &=& \sqrt{M_0^{\prime2}-(m^\prime_1-m_2)^2},\nonumber\\
 e^{(\prime)}_i  &=&\sqrt{m^{(\prime)2}_i+p^{\prime2}_\bot+p^{\prime2}_z},\nonumber\\
 p^\prime_z &=& \frac{x_2 M^\prime_0}{2}-\frac{m_2^2+p^{\prime2}_\bot}{2 x_2 M^\prime_0},\nonumber
 \end{eqnarray}
where $e_i^{(\prime)}$ is the energy of the quark and antiquark.
$M_0^\prime$ can be interpreted as the kinematic invariant mass in
the meson system.

For the transition of $B_{(s)}$ into a vector charmed/charmed-strange meson via the semileptonic decays, the transition matrix elements can be expressed by the form factors, i.e.,
\begin{eqnarray}
 \langle V(P^{\prime\prime},\varepsilon^{\prime\prime})|V_\mu|B_{(s)}(P^\prime)\rangle
          &=&\epsilon_{\mu\nu\alpha \beta}\,\varepsilon^{\prime\prime*\nu}P^\alpha q^\beta\, g({q^2}),
 \\
\langle V(P^{\prime\prime},\varepsilon^{\prime\prime})|A_\mu|B_{(s)}(P^\prime)\rangle
          &=&-i\bigg\{\varepsilon_\mu^{\prime\prime*} f({q^2})
              +\varepsilon^{*\prime\prime}\cdot P \Big[P_\mu a_+({q^2})
 \nonumber \\&&
              +q_\mu a_-({q^2})\Big]\bigg\}.
\end{eqnarray}
For the transition of $B_{(s)}$ into a $P-$wave charmed/charmed-strange meson via the semileptonic decays, the transition matrix elements are
\begin{eqnarray}
 \langle S(P^{\prime\prime})|A_\mu|B_{(s)}(P^\prime)\rangle &=& i\Big[u_+(q^2)P_\mu+u_-(q^2)q_\mu
 \Big], \\
 \langle A(P^{\prime\prime},\varepsilon^{\prime\prime})|V_\mu|B_{(s)}(P^\prime)\rangle
           &=& i\Big\{\ell(q^2)\varepsilon_\mu^{\prime\prime*}+\varepsilon^{\prime\prime*}\cdot
                  P[P_\mu c_+(q^2)
 \nonumber \\&&
                  +q_\mu c_-(q^2)]\Big\},\label{aaa}
  \\
 \langle A(P^{\prime\prime},\varepsilon^{\prime\prime})|A_\mu|B_{(s)}(P^\prime)\rangle
           &=& -q(q^2)\epsilon_{\mu\nu\alpha\beta}\varepsilon^{\prime\prime*\nu}P^\alpha
q^\beta,
 \\
 \langle T(P^{\prime\prime},\varepsilon^{\prime\prime})|V_\mu|B_{(s)}(P^\prime)\rangle
           &=& h(q^2)\epsilon_{\mu\nu\alpha\beta}\varepsilon^{\prime\prime*\nu\lambda}
                    P_\lambda P^\alpha q^\beta,
  \\
 \langle T(P^{\prime\prime},\varepsilon^{\prime\prime})|A_\mu|B_{(s)}(P^\prime)\rangle
           &=& -i\Big\{k(q^2)\varepsilon^{\prime\prime*}_{\mu\nu}P^{\nu}+
                \varepsilon^{\prime\prime*}_{\alpha\beta}P^{\alpha} P^{\beta}
                [P_\mu b_+(q^2)
 \nonumber \\&&
                +q_\mu b_-(q^2)]\Big\}.
\label{formfactordef}
\end{eqnarray}
 The corresponding form factors can be found in Refs. \cite{Jaus:1999zv,Cheng:2003sm}, which include
\begin{eqnarray}
 g(q^2)&=&-\frac{N_c}{16\pi^3}\int dx_2 d^2 p^\prime_\bot
           \frac{2 h^\prime_P h^{\prime\prime}_V}{x_2 \hat N^\prime_1 \hat N^{\prime\prime}_1}
           \Bigg\{x_2 m_1^\prime+x_1 m_2\nonumber\\&&+(m_1^\prime-m_1^{\prime\prime})
           \frac{p^\prime_\bot\cdot q_\bot}{q^2}
           +\frac{2}{w^{\prime\prime}_V}\left[p^{\prime2}_\bot+\frac{(p^\prime_\bot\cdot
            q_\bot)^2}{q^2}\right]
           \Bigg\},\nonumber\\
  \label{eq:formfactors1}
  \end{eqnarray}
  \begin{eqnarray}
  f(q^2)&=&\frac{N_c}{16\pi^3}\int dx_2 d^2 p^\prime_\bot
            \frac{ h^\prime_P h^{\prime\prime}_V}{x_2 \hat N^\prime_1 \hat N^{\prime\prime}_1}
            \Bigg\{2x_1(m_2-m_1^\prime)
  \nonumber\\
            &&\times(M^{\prime2}_0+M^{{\prime\prime}2}_0)-4 x_1
            m_1^{\prime\prime} M^{\prime2}_0+2x_2 m_1^\prime q\cdot P+2 m_2 q^2
  \nonumber\\
         &&-2 x_1 m_2
           (M^{\prime2}+M^{{\prime\prime}2})+2(m_1^\prime-m_2)(m_1^\prime+m_1^{\prime\prime})^2
  \nonumber\\
           &&+8(m_1^\prime-m_2)\left[p^{\prime2}_\bot+\frac{(p^\prime_\bot\cdot
            q_\bot)^2}{q^2}\right]+2(m_1^\prime+m_1^{\prime\prime})
  \nonumber\\
         &&
           \times(q^2+q\cdot P)\frac{p^\prime_\bot\cdot q_\bot}{q^2}
           -4\frac{q^2 p^{\prime2}_\bot+(p^\prime_\bot\cdot q_\bot)^2}{q^2 w^{\prime\prime}_V}
            \bigg[2 x_1
 \nonumber\\
          &&\times(M^{\prime2}+M^{\prime2}_0)-q^2-q\cdot P-2(q^2+q\cdot P)\frac{p^\prime_\bot\cdot
            q_\bot}{q^2}
 \nonumber\\
         &&-2(m_1^\prime-m_1^{\prime\prime})(m_1^\prime-m_2)
            \bigg]\Bigg\},
   \end{eqnarray}
  \begin{eqnarray}
 a_+(q^2)&=&\frac{N_c}{16\pi^3}\int dx_2 d^2 p^\prime_\bot
            \frac{2 h^\prime_P h^{\prime\prime}_V}{x_2 \hat N^\prime_1 \hat N^{\prime\prime}_1}
            \Bigg\{(x_1-x_2)(x_2 m_1^\prime
  \nonumber\\
            &&+x_1 m_2)-[2x_1
            m_2+m_1^{\prime\prime}+(x_2-x_1)
            m_1^\prime]\frac{p^\prime_\bot\cdot q_\bot}{q^2}
  \nonumber\\
         &&-2\frac{x_2 q^2+p_\bot^\prime\cdot q_\bot}{x_2 q^2
            w^{\prime\prime}_V}\Big[p^\prime_\bot\cdot p^{\prime\prime}_\bot+(x_1 m_2+x_2 m_1^\prime)
  \nonumber\\
            &&\times(x_1 m_2-x_2
            m_1^{\prime\prime})\Big]\Bigg\},
\end{eqnarray}
\begin{eqnarray}
 a_-(q^2)&=&\frac{N_c}{16\pi^3}\int dx_2 d^2 p^\prime_\bot
            \frac{ h^\prime_P h^{\prime\prime}_V}{x_2 \hat N^\prime_1 \hat N^{\prime\prime}_1}
            \Bigg\{2(2x_1-3)(x_2 m_1^\prime
 \nonumber\\
            &&+x_1 m_2)-8(m_1^\prime-m_2)
            \left[\frac{p^{\prime2}_\bot}{q^2}+2\frac{(p^\prime_\bot\cdot q_\bot)^2}{q^4}\right]\nonumber\\
         &&-[(14-12 x_1) m_1^\prime -2 m_1^{\prime\prime}-(8-12 x_1) m_2]\frac{p^\prime_\bot\cdot q_\bot}{q^2}\nonumber\\
         &&+\frac{4}{w^{\prime\prime}_V}\bigg([M^{\prime2}+M^{\prime\prime2}-q^2+2(m_1^\prime-m_2)(m_1^{\prime\prime}+m_2)]
 \nonumber\\
                                   &&\times(A^{(2)}_3+A^{(2)}_4-A^{(1)}_2)+Z_2(3 A^{(1)}_2-2A^{(2)}_4-1)
 \nonumber\\
         &&+\frac{1}{2}[x_1(q^2+q\cdot P)
            -2 M^{\prime2}-2 p^\prime_\bot\cdot q_\bot -2 m_1^\prime
 \nonumber\\
            &&\times(m_1^{\prime\prime}+m_2)-2 m_2(m_1^\prime-m_2)](A^{(1)}_1+A^{(1)}_2-1)
 \nonumber\\
         &&+
         q\cdot P\Bigg[\frac{p^{\prime2}_\bot}{q^2}
         +\frac{(p^\prime_\bot\cdot q_\bot)^2}{q^4}\Bigg] (4A^{(1)}_2-3)\bigg)
            \Bigg\},\label{eq:formfactors4}
\end{eqnarray}
which are related to the form factors in Eqs. (\ref{eq:B TO_V1}) and (\ref{eq:B TO_V}), i.e.,
\begin{eqnarray}
 V^{B_{(s)}\to V}(q^2)&=&-(m_{B_{(s)}}+m_{V})\, g(q^2),\\
  A_1^{B_{(s)}\to V}(q^2)&=&-\frac{f(q^2)}{m_{B_{(s)}}+m_{V}},\\
 A_2^{B_{(s)}\to V}(q^2)&=&(m_{B_{(s)}}+m_{V})\, a_+(q^2),\\
 A_3^{B_{(s)}\to V}(q^2)-A_0^{B_{(s)}\to V}(q^2)&=&\frac{q^2}{2 m_{V}}\,
a_-(q^2).\label{eq:relation-vector}
\end{eqnarray}

Similarly, the expressions for the form factors for  the hadronic
matrix elements of $B_{(s)} \to{^3A}$ and $B_{(s)} \to{^1A}$ obtained by replacing $f(q^2)$,
$g(q^2)$, and $a_\pm(q^2)$ in Eqs.
(\ref{eq:formfactors1})-(\ref{eq:formfactors4})
\cite{Cheng:2003sm}, i.e.,
 \begin{eqnarray}
 \ell^{^A}(q^2)&=&f(q^2) \,\,\,{\rm with}\label{aa}\\ &&
                         (m_1^{\prime\prime}\to -m_1^{\prime\prime},\,h^{\prime\prime}_V\to h^{\prime\prime}_{^3A},
                         \,w^{\prime\prime}_V\to w^{\prime\prime}_{^A}),
 \nonumber\\
 q^{^A}(q^2)&=&g(q^2) \,\,\,{\rm with}\\ &&
                         (m_1^{\prime\prime}\to -m_1^{\prime\prime},\,h^{\prime\prime}_V\to h^{\prime\prime}_{^3A},
                         \,w^{\prime\prime}_V\to  w^{\prime\prime}_{^A}),
  \nonumber\\
 c_\pm^{^A}(q^2)&=&a_\pm(q^2) \,\,\,{\rm with}\nonumber\\ &&
                          (m_1^{\prime\prime}\to -m_1^{\prime\prime},\,h^{\prime\prime}_V\to h^{\prime\prime}_{^3A},
                          \,w^{\prime\prime}_V\to  w^{\prime\prime}_{^A}).
                          \label{eq:Aformfactor}
 \end{eqnarray}
It should be noticed that only the ${1}/{w^{\prime\prime}}$ term is left for the $^1A$ charmed meson. Then, the form factors  in Eqs. (\ref{eq:B TO A1}) and (\ref{eq:B TO A}) have the relations
 \begin{eqnarray}
 A^{B_{(s)}\to A}(q^2)&=&-(m_{B_{(s)}}-m_A)\, q^A(q^2), \label{ABSWFF}\\
 V^{B_{(s)}\to A}_1(q^2)&=&-\frac{\ell^A(q^2)}{m_{B_{(s)}}-m_A},\label{k1}
\\
 V^{B_{(s)}\to A}_2(q^2)&=&(m_{B_{(s)}}-m_A)\, c_+^A(q^2), \\
 V_3^{B_{(s)}\to A}(q^2)-V_0^{B_{(s)}\to A}(q^2)&=&\frac{q^2}{2 m_A}\,
 c_-^A(q^2) \label{V0BSWFF}.
 \end{eqnarray}

Analogously, the form factors for the hadronic matrix element of  $B_{(s)} \to S$ are
\begin{eqnarray}
 u_+(q^2)&=&\frac{N_c}{16\pi^3}\int dx_2 d^2p^\prime_\bot
            \frac{h^\prime_P h^{\prime\prime}_S}{x_2 \hat N_1^\prime \hat N^{\prime\prime}_1}
            \Big[-x_1 (M_0^{\prime2}+M_0^{\prime\prime2}) \nonumber\\&&
            -x_2 q^2+x_2(m_1^\prime+m_1^{\prime\prime})^2 +x_1(m_1^\prime-m_2)^2 \nonumber\\&&
            +x_1(m_1^{\prime\prime}+m_2)^2\Big],
\end{eqnarray}
\begin{eqnarray}
 u_-(q^2)&=&\frac{N_c}{16\pi^3}\int dx_2 d^2p^\prime_\bot
            \frac{2h^\prime_P h^{\prime\prime}_S}{x_2 \hat N_1^\prime \hat N^{\prime\prime}_1}
            \Bigg\{ x_1 x_2 M^{\prime2}+p_\bot^{\prime2} \nonumber\\&&
            +m_1^\prime m_2
                  +(m_1^{\prime\prime}+m_2)(x_2 m_1^\prime+x_1 m_2)
\nonumber\\
         && -2\frac{q\cdot P}{q^2}\left(p^{\prime2}_\bot+2\frac{(p^\prime_\bot\cdot q_\bot)^2}{q^2}\right)
                  -2\frac{(p^\prime_\bot\cdot q_\bot)^2}{q^2}\nonumber\\&&
                  +\frac{p^\prime_\bot\cdot q_\bot}{q^2}
                  \Big[M^{\prime\prime2}-x_2(q^2+q\cdot P)-(x_2-x_1) M^{\prime2}\nonumber\\&&
         +2 x_1 M_0^{\prime2}-2(m_1^\prime-m_2)(m_1^\prime-m_1^{\prime\prime})\Big]
           \Bigg\},
 \label{eq:uformfactor}
\end{eqnarray}
which are related to the form factors in Eq. (\ref{eq:B TO_S}),
\begin{eqnarray}
 F^{B_{(s)}\to S}_1(q^2)&=&-u_+(q^2), \label{F1BSWFF}\\
                 F^{B_{(s)}\to S}_0(q^2)&=&-u_+(q^2)-\frac{q^2}{q\cdot P} u_-(q^2) \label{F0BSWFF}.
\end{eqnarray}

For the hadronic matrix element of $B_{(s)} \to T$, the form factors in Eqs. (\ref{eq:B TO T1}) and (\ref{eq:B TO T}) can be written as
\begin{eqnarray}
&& h(q^2)\nonumber\\&&=-g(q^2)\Big|_{h^{\prime\prime}_V\to h^{\prime\prime}_T}+\frac{N_c}{16\pi^3}\int dx_2 d^2 p^\prime_\bot
            \frac{2 h^\prime_P h^{\prime\prime}_T}{x_2 \hat N^\prime_1 \hat N^{\prime\prime}_1}
            \Bigg[(m_1^\prime
\nonumber\\
         &&\quad-m_1^{\prime\prime})(A^{(2)}_3+A^{(2)}_4)
          +(m_1^{\prime\prime}+m_1^\prime-2 m_2)(A^{(2)}_2+A^{(2)}_3)\nonumber\\&&\quad
                   -m_1^\prime (A^{(1)}_1+A^{(1)}_2)
                   +\frac{2}{w^{\prime\prime}_V}(2 A^{(3)}_1+2 A^{(3)}_2-A^{(2)}_1)
            \Bigg],
 \end{eqnarray}
 \begin{eqnarray}
&& k(q^2)\nonumber\\&&=-f(q^2)\Big|_{h^{\prime\prime}_V\to h^{\prime\prime}_T}+\frac{N_c}{16\pi^3}\int dx_2 d^2 p^\prime_\bot
            \frac{ h^\prime_P h^{\prime\prime}_T}{x_2 \hat N^\prime_1 \hat N^{\prime\prime}_1}
            \Bigg\{2(A^{(1)}_1
 \nonumber\\
           &&\quad+A^{(1)}_2)[m_2(q^2-\hat N_1^\prime-\hat N^{\prime\prime}_1-m_1^{\prime2}-m_1^{\prime\prime2})
 \nonumber\\
           &&   \quad   -m_1^\prime (M^{\prime\prime2}-\hat N_1^{\prime\prime}-m_1^{\prime\prime2}-m_2^2)
 \nonumber\\
           &&\quad -m^{\prime\prime}_1(M^{\prime2}-\hat N^\prime_1-m_1^{\prime2}-m_2^2)-2 m_1^\prime m_1^{\prime\prime}
                   m_2]
       +2(m_1^\prime
 \nonumber\\
           &&\quad+m_1^{\prime\prime})\Bigg(A^{(1)}_2 Z_2+\frac{q\cdot P}{q^2} A^{(2)}_1\Bigg)+16(m_2-m_1^\prime)(A^{(3)}_1 \nonumber\\
          &&                \quad   +A^{(3)}_2)+4(2 m_1^\prime-m_1^{\prime\prime}-m_2) A^{(2)}_1
       +\frac{4}{w^{\prime\prime}_V}\Bigg([M^{\prime2}+M^{\prime\prime2}
 \nonumber\\
           && \quad-q^2(2 A^{(3)}_1+2
                                           A^{(3)}_2-A^{(2)}_1)
                                           +2(m_1^\prime-m_2)(m_1^{\prime\prime}+m_2)] \nonumber\\
          &&      \quad                                     \times
                                -4\Bigg[A^{(3)}_2 Z_2+\frac{q\cdot P}{3q^2}\Big(A^{(2)}_1\Big)^2
                                              \Bigg]
                                            +2A^{(2)}_1 Z_2
                                     \Bigg)
            \Bigg\},
      \end{eqnarray}
 \begin{eqnarray}
 && b_+(q^2)\nonumber\\&&=-a_+(q^2)\Big|_{h^{\prime\prime}_V\to h^{\prime\prime}_T}+\frac{N_c}{16\pi^3}\int dx_2 d^2 p^\prime_\bot
            \frac{ h^\prime_P h^{\prime\prime}_T}{x_2 \hat N^\prime_1 \hat N^{\prime\prime}_1}
            \Bigg\{8(m_2
  \nonumber\\
          &&\quad-m_1^\prime)(A^{(3)}_3+2A^{(3)}_4+A^{(3)}_5)
       -2m_1^\prime (A^{(1)}_1+A^{(1)}_2)
  \nonumber\\
          &&          \quad        +4(2 m_1^\prime-m_1^{\prime\prime}-m_2)
                   (A^{(2)}_2+A^{(2)}_3)
                   +2(m_1^\prime+m_1^{\prime\prime})
  \nonumber\\
          &&       \quad   \times(A^{(2)}_2+2A^{(2)}_3+A^{(2)}_4)
       +\frac{2}{w^{\prime\prime}_V}\Bigg[2[M^{\prime2}+M^{\prime\prime2}-q^2
  \nonumber\\
          &&       \quad                              +2(m_1^\prime-m_2)(m_1^{\prime\prime}+m_2)](A^{(3)}_3+2 A^{(3)}_4+A^{(3)}_5-A^{(2)}_2
  \nonumber\\
          &&             \quad                       -A^{(2)}_3)
                                          +[q^2-\hat N_1^\prime-\hat N_1^{\prime\prime}-(m_1^\prime+m_1^{\prime\prime})^2]
  \nonumber\\
          &&           \quad                  \times(A^{(2)}_2+2A^{(2)}_3+A^{(2)}_4-A^{(1)}_1-A^{(1)}_2)
                                     \Bigg]
            \Bigg\},
             \end{eqnarray}
 \begin{eqnarray}
   &&b_-(q^2)\nonumber\\&&=-a_-(q^2)\Big|_{h^{\prime\prime}_V\to h^{\prime\prime}_T}+\frac{N_c}{16\pi^3}\int dx_2 d^2 p^\prime_\bot
            \frac{ h^\prime_P h^{\prime\prime}_T}{x_2 \hat N^\prime_1 \hat N^{\prime\prime}_1}
            \Bigg\{8(m_2
  \nonumber\\
          &&  \quad   -m_1^\prime)(A^{(3)}_4+2A^{(3)}_5+A^{(3)}_6)
       -6m_1^\prime (A^{(1)}_1+A^{(1)}_2)
                   \nonumber\\
          && \quad+4(2 m_1^\prime-m_1^{\prime\prime}-m_2)
                   (A^{(2)}_3+A^{(2)}_4)
       +2(3m_1^\prime+m_1^{\prime\prime}
 \nonumber\\
          &&\quad-2m_2)(A^{(2)}_2+2A^{(2)}_3+A^{(2)}_4)  +\frac{2}{w^{\prime\prime}_V}\Bigg[2[M^{\prime2}+M^{\prime\prime2}
 \nonumber\\
          &&    \quad -q^2
                                           +2(m_1^\prime-m_2)(m_1^{\prime\prime}+m_2)]
                                           (A^{(3)}_4+2 A^{(3)}_5+A^{(3)}_6
 \nonumber\\
          &&             \quad            -A^{(2)}_3-A^{(2)}_4)
                                           +2Z_2(3A^{(2)}_4-2A^{(3)}_6-A^{(1)}_2)
 \nonumber\\
          &&
                             \quad              +2\frac{q\cdot P}{q^2}
                                             \Big(6 A^{(1)}_2 A^{(2)}_1-6A^{(1)}_2A^{(3)}_2
                                             +\frac{2}{q^2}\big(A^{(2)}_1\big)^2-A^{(2)}_1\Big)
 \nonumber\\
          &&         \quad                +[q^2-2M^{\prime2}+\hat N_1^\prime-\hat N_1^{\prime\prime}
                                                -(m_1^\prime+m_1^{\prime\prime})^2+2(m_1^\prime-m_2)^2]
 \nonumber\\
          &&          \quad                      \times(A^{(2)}_2+2A^{(2)}_3+A^{(2)}_4-A^{(1)}_1-A^{(1)}_2)
                                     \Bigg]
            \Bigg\},
 \label{eq:formfactorT}
\end{eqnarray}
where $g(q^2)$, $f(q^2)$, $a_+$, and $a_-$ are given in Eqs.
(\ref{eq:formfactors1})-(\ref{eq:formfactors4}), respectively.
With the above results, one can redefine the form factors as
\begin{eqnarray}
 A^{B_{(s)}\to T}&=&-(m_{B_{(s)}}-m_T)h(q^2),
 \end{eqnarray}
 \begin{eqnarray}
  V_1^{B_{(s)}\to T}&=&-\frac{k(q^2)}{m_{B_{(s)}}-m_T},
   \end{eqnarray}
 \begin{eqnarray}
  V_2^{B_{(s)} \to T}&=& (m_{B_{(s)}}-m_T)b_+(q^2),
   \end{eqnarray}
 \begin{eqnarray}
 V_0^{B_{(s)}\to T}(q^2)&=&\frac{m_{B_{(s)}}-m_T}{2m_T} V_1^{B_{(s)}\to T}(q^2)\nonumber\\&&-\frac{m_{B_{(s)}}+m_T}{2m_T}
 V_2^{B_{(s)}\to  T}(q^2)\nonumber\\&&
 -\frac{q^2}{2m_T}b_-(q^2).\label{B TO T}
\end{eqnarray}
Although these redefined  form factors of the hadronic matrix element of $B_{(s)} \to T$ are not  dimensionless,
this treatment is convenient for calculating the corresponding decay width \cite{Wang:2009mi}, which is given in Sec. \ref{sec3}.

The concrete expressions for $A_1^{(1)}$, $A_2^{(1)}$, $A_1^{(2)}$, $A_2^{(2)}$, $A_3^{(2)}$, $A_4^{(2)}$, $A_1^{(3)}$, $A_2^{(3)}$, $A_3^{(3)}$, $A_4^{(3)}$, $A_5^{(3)}$, $A_6^{(3)}$, and $Z_2$ are as follows:
\begin{eqnarray}
 A^{(1)}_1&=&\frac{x_1}{2},
 \quad
 A^{(1)}_2=A^{(1)}_1-\frac{p^\prime_\bot\cdot q_\bot}{q^2},
 \quad
 \nonumber
    \end{eqnarray}
  \begin{eqnarray}
 Z_2&=&\hat N_1^\prime+m_1^{\prime2}-m_2^2+(1-2x_1)M^{\prime2}
 +(q^2+q\cdot P)\frac{p^\prime_\bot\cdot q_\bot}{q^2}, \nonumber
 \end{eqnarray}
  \begin{eqnarray}
 A^{(2)}_1&=&-p^{\prime2}_\bot-\frac{(p^\prime_\bot\cdot q_\bot)^2}{q^2},
 \quad
 A^{(2)}_2=\big(A^{(1)}_1\big)^2,
 \quad
 A^{(2)}_3=A^{(1)}_1 A^{(1)}_2,
 \nonumber
    \end{eqnarray}
  \begin{eqnarray}
 A^{(2)}_4&=&\big(A^{(1)}_2\big)^2-\frac{1}{q^2}A^{(2)}_1,
 \quad
 A^{(3)}_1=A^{(1)}_1 A^{(2)}_1,
 \quad
 A^{(3)}_2=A^{(1)}_2 A^{(2)}_1,
  \nonumber
     \end{eqnarray}
  \begin{eqnarray}
 A^{(3)}_3&=&A^{(1)}_1 A^{(2)}_2,
 \quad
 A^{(3)}_4=A^{(1)}_2 A^{(2)}_2,
 \quad
 A^{(3)}_5=A^{(1)}_1 A^{(2)}_4,
  \nonumber
     \end{eqnarray}
  \begin{eqnarray}
 A^{(3)}_6&=&A^{(1)}_2 A^{(2)}_4-\frac{2}{q^2}A^{(1)}_2 A^{(2)}_1.\nonumber
\end{eqnarray}

The concrete expressions of the functions in Eqs. (\ref{HQSR1})-(\ref{HQSR33}) are as follows:
\begin{eqnarray}
 && \tau_{1/2}^u={1\over 4\sqrt{m_{B_{(s)}}m_S}}\Big[(m_{B_{(s)}}-m_S)u_+(q^2) \nonumber\\&&\qquad\quad+(m_{B_{(s)}}+m_S)u_-(q^2)\Big],  \label{f1}\\
 && \tau_{1/2}^\ell= {1\over 2\sqrt{m_{B_{(s)}}m_{A^{1/2}}}}{\ell^{1/2}(q^2)\over \omega-1}, \label{f3} \\
 && \tau_{1/2}^q= \sqrt{m_{B_{(s)}}m_{A^{1/2}}}\,q_{1/2}(q^2) , \\
 && \tau_{1/2}^c= -{\sqrt{m_{B_{(s)}}m_{A^{1/2}}}\over 2}\Big(c^{1/2}_+(q^2)  -c^{1/2}_-(q^2)\Big),\\
&&\tau_{3/2}^\ell= -\sqrt{2\over m_{B_{(s)}}m_{A^{3/2}}}\,{\ell^{3/2}(q^2)\over \omega^2-1}, \label{f4} \\
 &&\tau_{3/2}^{c+}=-{1\over 3}\sqrt{2m_{B_{(s)}}^3
 \over m_{A^{3/2}}}\left(c^{3/2}_+(q^2)+c^{3/2}_-(q^2)\right),
 \\
 &&\tau_{3/2}^{c-}= -\sqrt{2m_{B_{(s)}}^3\over m_{A^{3/2}}}\,{c^{3/2}_+(q^2)-c^{3/2}_-(q^2)\over
 \omega-2},  \\
 &&\tau_{3/2}^h= 2\,\sqrt{m_{B_{(s)}}^3m_T\over 3}\,h(q^2) , \\
 &&\tau_{3/2}^k= \sqrt{m_{B_{(s)}}\over 3m_T}{k(q^2)\over 1+\omega} ,  \\
 &&\tau_{3/2}^q=-{2\sqrt{2}\over 1+\omega}{\sqrt{m_{B_{(s)}}m_{A^{3/2}}}}~q_{3/2}(q^2), \\
 &&\tau_{3/2}^b= - \sqrt{m_{B_{(s)}}^3m_T\over
 3}\left(b_+(q^2)-b_-(q^2)\right).\label{f2}
\end{eqnarray}

\end{document}